\documentclass[prb,aps,showpacs,preprintnumbers,amsmath,amssymb,twocolumn,superscriptaddress]{revtex4-2}

\usepackage[utf8]{inputenc}
\usepackage{graphicx,pbox,lineno}
\graphicspath{{./Figuras/}}
\usepackage[dvipsnames]{xcolor}
\usepackage[normalem]{ulem}
\allowdisplaybreaks
\usepackage[colorlinks, citecolor={blue!50!black}, urlcolor={blue!50!black}, linkcolor={red!50!black}]{hyperref}
\usepackage[caption=false]{subfig}
\usepackage{units}
\usepackage{mathtools}
\usepackage{amsmath}
\usepackage{amssymb}
\usepackage{bbold}

\begin{document}


\title{Strain-engineering the topological type-II Dirac semimetal NiTe$_2$}
\author{P. P. Ferreira}
\email[Corresponding author: ]{pedroferreira@usp.br}
\affiliation{Computational Materials Science Group (ComputEEL/MatSci), Escola de Engenharia de Lorena, Universidade de S\~ao Paulo, Materials Engineering Department, Lorena, Brazil}
\author{A. L. R. Manesco}
\email[Corresponding author: ]{antoniolrm@usp.br}
\affiliation{Computational Materials Science Group (ComputEEL/MatSci), Escola de Engenharia de Lorena, Universidade de S\~ao Paulo, Materials Engineering Department, Lorena, Brazil}
\affiliation{Kavli Institute of Nanoscience, Delft University of Technology, Delft, The Netherlands}
\author{T. T. Dorini}
\affiliation{Universit\'e de Lorraine, CNRS, IJL, Nancy, France}
\author{L. E. Correa}
\author{{G. Weber}}
\author{A. J. S. Machado}
\author{L. T. F. Eleno}
\email[Corresponding author: ]{luizeleno@usp.br}
\affiliation{Computational Materials Science Group (ComputEEL/MatSci), Escola de Engenharia de Lorena, Universidade de S\~ao Paulo, Materials Engineering Department, Lorena, Brazil}

\begin{abstract}
In the present work, we investigated the electronic and elastic properties in equilibrium and under strain of the type-II Dirac semimetal NiTe$_2$ using density functional theory (DFT). Our results demonstrate the tunability of Dirac nodes' energy and momentum with strain and that it is possible to bring them closer to the Fermi level, while other metallic bands are supressed. We also derive a minimal 4-band effective model for the Dirac cones which accounts for the aforementioned strain effects by means of lattice regularization, providing an inexpensive way for further theoretical investigations and easy comparison with experiments. On an equal footing, we propose the static control of the electronic structure by intercalating alkali species into the van der Waals gap, resulting in the same effects obtained by strain-engineering and removing the requirement of in situ strain. Finally, evaluating the wavefunction's symmetry evolution as the lattice is deformed, we discuss possible consequences, such as Liftshitz transitions and the coexistence of type-I and type-II Dirac cones, thus motivating future investigations.
\end{abstract}

\date{\today}


\maketitle

\section{Introduction}

The existence of quasiparticle excitations with no counterpart in high energy physics became relevant, not only due to mere scientific interest, but also for the possibility of using their properties as building blocks for new electronic devices. From the description of the quantum spin Hall effect in graphene by Kane and Mele \cite{kane2005a, kane2005b} and the first realization of three-dimensional topological band insulators \cite{fu2007a, fu2007b, hsieh2008, hsieh2009, xia2009, zhang2009, chen2009} to the proposal of topological metallic states \cite{vafek2014, burkov2016a, weng2016}, as Weyl and Dirac semimetals \cite{armitage2018}, the existence of such novel quasiparticles has drawn a heightened interest in the last few years. Among their properties, one could highlight the ultrahigh electronic mobility and conductivity \cite{shekhar2015, liang2015, zhao2015, xiong2015}, negative/giant magnetoresistence \cite{huang2015a, li2016, gao2017}, chiral anomaly \cite{zyuzin2012, parameswaran2014, zhang2016, burkov2016b}, and quantum anomalous Hall effect \cite{haldane2004, xu2011, weng2015}.

The simplest example of a Dirac material is graphene, for which valence and conduction bands touch at discrete points in the first Brillouin zone and disperse linearly in all momentum directions \cite{PhysRev.71.622, RevModPhys.81.109}. In three-dimensions, twofold (Weyl) or fourfold (Dirac) symmetry-protected degenerate points host bulk massless fermionic quasiparticle excitations and surface spin-textures, robust against pertubations \cite{wan2011, liu2014, yi2014, huang2015b, xu2015}. Thus, breaking the twofold degeneracy ensured by either inversion- or time-reversal symmetries, a Dirac cone will decouple into a pair of opposite-chirallity Weyl fermions \cite{wang2012, zyuzin2012b, okugawa2014}.

\begin{figure}[b]
	\includegraphics[width=\columnwidth]{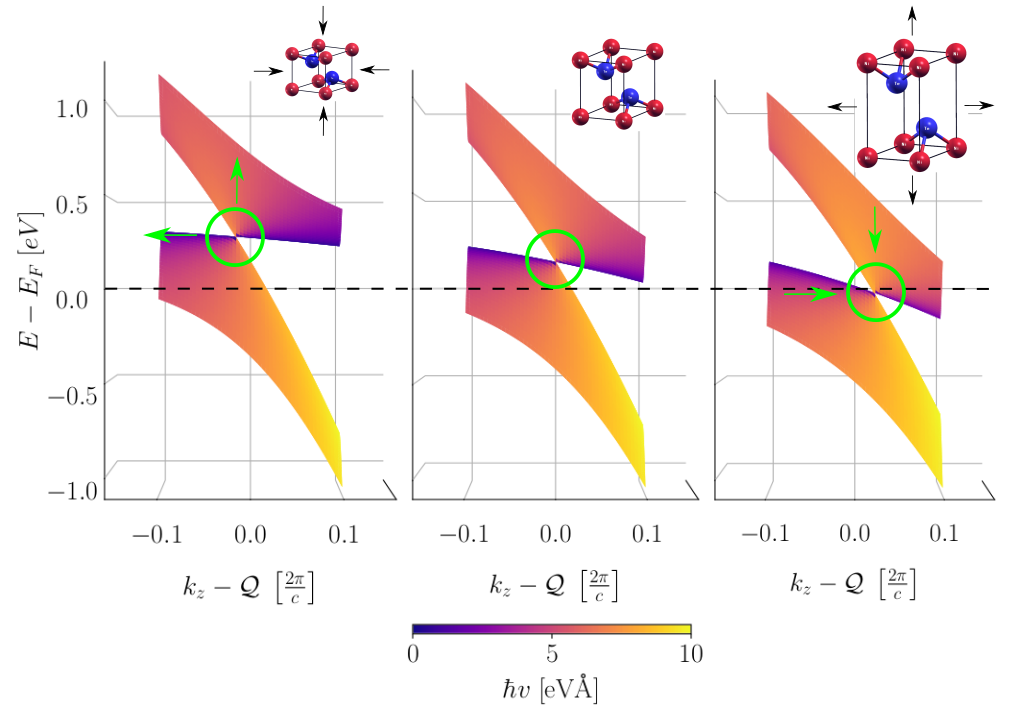}
	\caption{Dirac cone tunability as a function of isostatic pressure according to our effective model. The middle plot shows the Dirac cone for the ground-state structure.
		The left and right plots show the Dirac cone under compressive and tensile strain, respectively. 
		Under compressive strain the Dirac cone energy increases while the Dirac node's $k_z$-component decrease, while the opposite happens under tensile strain (indicated by the green arrows in both cases). It is possible to see that the Dirac cone crosses the Fermi energy level under tensile strain, making NiTe$_2$ an exciting platform for electronic transport experiments.}
	\label{fig:dirac_tunability}
\end{figure}

Different from their high energy physics counterparts, massless quasiparticles in solids move at the Fermi velocity, thus their dispersion is not effectively constrained by Lorentz invariance. Therefore, the energy-momentum dispersion explicitly depends on its direction in momentum space \cite{soluyanov2015}. The resulting tilted cones are the characteristic signature of the so-called type-II Weyl and Dirac semimetals \cite{deng2016, koepernik2016, chang2016, zhang2017, noh2017, yan2017}, and show, for instance, anisotropic transport and magnetoresistance properties \cite{wang2016, chen2016, kumar2017, lai2018}, in contrast to type-I materials.

In this context, the transition metal dichalcogenide NiTe$_2$ was recently rediscovered as a type-II Dirac semimetal \cite{xu2018, de2018, zheng2020}. Transport measurements revealed a non-saturating linear magnetoresistance and quantum oscillations confirmed the existence of a nontrivial Berry phase for the light mass carriers \cite{xu2018}. The existence of topological surface states with chiral spin-texture over a wide range of energies was supported by spin- and angle-resolved photoemission spectroscopy \cite{ghosh2019}. Additionally, superconductivity was observed in NiTe$_2$ under pressure \cite{li2019} and with the intercalation of Ti into the van der Waals gap (the space between two adjacent chalcogenide layers) \cite{de2018}, and was also predicted in atomically thin systems \cite{zheng2019}. Moreover, the energy position of its Dirac node, closer to the Fermi level when compared with similar systems \cite{PhysRevB.94.121117, zhang2017, xiao2017}, combined with accessible high-quality single-crystals \cite{monteiro2017, zhao2018, liu2019} substantiate the interest on the material.

In the present work, we study, from first-principles calculations, strain effects on the electronic structure of NiTe$_2$. Our results show that it is possible to systematically tune both the energy relative to the Fermi level and the point in the Brillouin zone where the type-II Dirac cone is located, as illustrated in Fig. \ref{fig:dirac_tunability}, which summarizes some of our findings. While the the energy of the cone is relevant for enhancing/supressing the effects of states with massless dispersion on transport properties, its position in the Brillouin zone provides a route to create artificial magnetic fields in this material \cite{PhysRevX.6.041046, PhysRevX.6.041021}. Additionally, the evolution of the irreducible representations for the electronic states under strain shows the appearance of a type-I Dirac cone in the same pair of bands as the type-II Dirac cone, establishing a hybrid pseudo-relativistic topological phase. Finally, we also demonstrate that alkali metal intercalation into the van der Waals gap acts effectively as a static chemical-pressure source inside the crystal-structure, simulating the effects obtained by strain-engineering.

The manuscript is organized as follows. Sec. \ref{sec:methods} describes the computational methods and numerical parameters used in the first-principle electronic-structure calculations. In Sec. \ref{sec:properties-in-equilibrium}, we present a comprehensive investigation of the electronic and elastic properties of the ground-state structure. Sec. \ref{sec:strain} shows the key results related to the strain-engineering of the electronic states of NiTe$_2$ in the vicinity of the Fermi level. Finally, Sec. \ref{sec:minimal-model} is devoted to constructing a 4-band low-energy effective model for the type-II Dirac cone.

\section{Computational Methods}

\label{sec:methods}

First-principles electronic-structure calculations were carried out in the framework of the Density Functional Theory (DFT) within the Kohn-Sham scheme \cite{hohenberg1964, kohn1965}, using the pseudopotential approach as implemented in \emph{Quantum Espresso} \cite{giannozzi2009, giannozzi2017} and auxiliary post-processing tools \cite{kokalj1999, kawamura2019}. The calculations were performed using a series of different approximations for the exchange and correlation (XC) functional, within its relativistic and non-relativistic forms: local-density-approximation (LDA) of Perdew-Zhang (PZ) \cite{perdew1981} and the generalized-gradient-approximation with the Perdew-Burke-Ernzerhof (PBE) \cite{perdew1996} parametrization and its modified version, known as PBEsol \cite{perdew2008}; as well as non-local functionals, including the van der der Waals interactions \cite{tran2019}, namely vdW-DF \cite{dion2004, dion2005}, optB86b-vdW \cite{klimevs2009}, and optB88-vdW \cite{klimevs2011}. To guarantee the energy eigenvalues convergence from the Kohn-Sham self-consistent solution, we have adopted a wavefunction energy cut-off of \unit[260]{Ry} and a sampling of $16\times16\times8$ $k$-points in the first Brillouin zone according to the Monkhorst-Pack scheme \cite{monkhorst1976}. To compute the electronic properties, a denser $k$-mesh grid was considered, with $32\times32\times16$ $k$-points. All lattice parameters and internal degrees of freedom were fully relaxed in order to reach a ground-state convergence of \unit[10$^{-6}$]{Ry} in total energy and \unit[10$^{-4}$]{Ry/a$_0$} (a$_0$ $\approx$ 0.529\,\AA) for forces acting on the nuclei.

The full second-order elastic stiffness tensor was obtained from a set of deformations imposed on the underformed reference ground-state ($\eta=0$) structures, as implemented in the \emph{ElaStic} tool \cite{golesorkhtabar2013}. To obtain the six independent second-order elastic constants of the trigonal symmetry, we have used six different types of deformation, with 15 distorted structures each and strain intensities in the range $-0.05 \le \eta \le +0.05$. The macroscopic mechanical moduli and their crystallographic-orientation dependence was derived from the stiffness tensor \cite{ferreira2018}. Details on this topic are provided in App. \ref{app:mech}.

Finally, the effective model was constructed using Qsymm \cite{Varjas_2018}. We found a family of hamiltonians up to second order in momentum, satisfying the same set of discrete symmetries as NiTe$_2$ and restricted to the orbitals forming the type-II Dirac cone. The strain-dependency was implemented using a lattice regularization scheme \cite{PhysRevX.6.041021}. All free parameters were then fitted with DFT data.

All code and data used to prepare this manuscript is freely available on the Zenodo repository \cite{antonio_manesco_2020_4279841}.

\section{Ground-state properties}
\label{sec:properties-in-equilibrium}

\subsection{Elastic properties}

\begin{figure}[t]
	\centering
		\subfloat[][]{\includegraphics[width=.3\columnwidth]{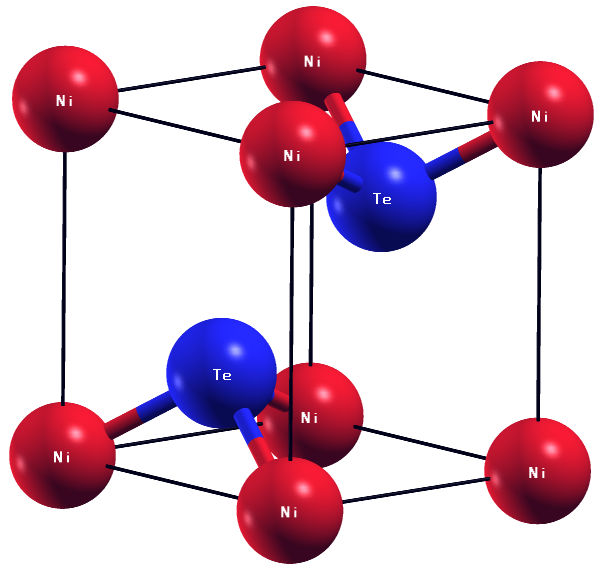}}
		\subfloat[][]{\includegraphics[width=.45\columnwidth]{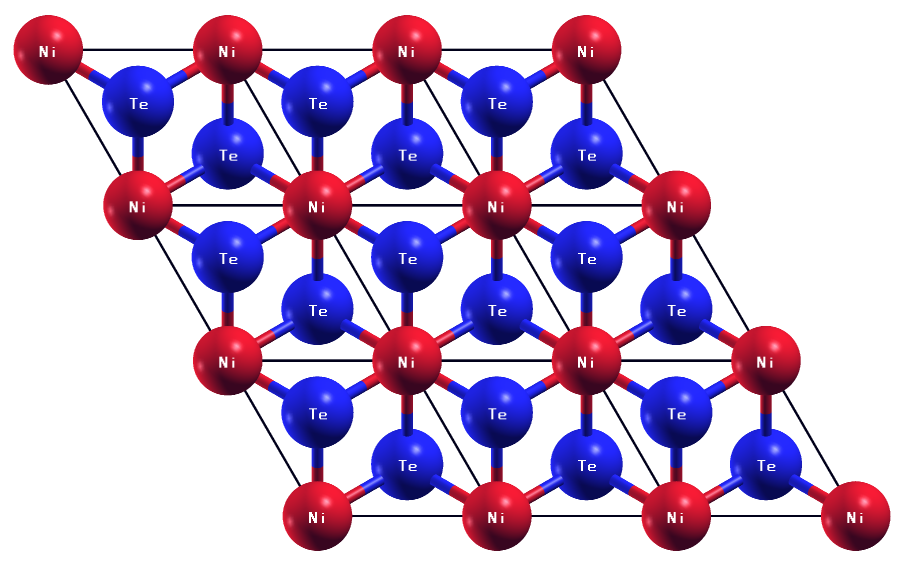}}\\
		\subfloat[][]{\includegraphics[width=.55\columnwidth]{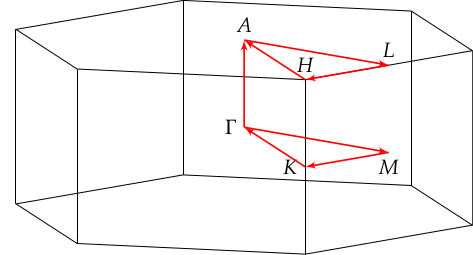}}
	\caption{(a) Trigonal unit cell of NiTe$_2$ alongside (b) a top view of a  $3\times 3\times 1$ supercell (blue: Te; red: Ni atoms). (c) First Brillouin zone of NiTe$_2$, with the path along high-symmetry points used to plot the dispersion curves \cite{setyawan2010}.}
	\label{fig:NiTe2}
\end{figure}

NiTe$_2$ is a layered compound that crystallizes in a trigonal centrosymmetric structure within CdI$_2$-prototype (space-group $P\bar{3}m1$, no. 164), as shown in Fig. \ref{fig:NiTe2}. A Ni layer is sandwiched between two Te layers, with the stacking of adjacent Te layers mediated by weak van der Waals interactions \cite{manzeli2017}. The optimized cell parameters and the relaxed Te-position degree of freedom are presented in Table \ref{tab:NiTe2-lat-par}. Regardless of the choice of the exchange and correlation functional for the Kohn-Sham Hamiltonian, the relative error, compared with the experimental crystallographic data available, for both calculated cell parameters and atomic positions does not exceed 3\% in our calculations. However, we have observed that spin-orbit coupling (SOC) effects are more expressive over the electronic energy dispersion than van der Waals interactions between adjacent layers. Thus, all the results presented in this manuscript, except when explicitly mentioned, correspond to the PBE parametrization including SOC effects.

\begin{table}[b]
	\caption{NiTe$_2$ fully optimized cell-parameters ($a$ and $c$) and atomic position degree of freedom of Te atoms ($z_\text{Te}$) using different XC functionals.}
	\label{tab:NiTe2-lat-par}
	\centering
	\begin{tabular}{lcccc}
	    \toprule
		XC functional& $a$ (\AA) & $c$ (\AA) & $V$ (\AA$^3$) & $z_\text{Te}$\\
		\hline
		PZ & 3.814 & 5.105 & 64.31 & 0.2522 \\[2mm]
		PZ+SOC & 3.797 & 5.186 & 64.75 & 0.2511 \\[2mm]
		PBE & 3.894 & 5.372 & 70.54 & 0.2442 \\[2mm]
		PBE+SOC & 3.897 & 5.377 & 70.72 & 0.2452 \\[2mm]
		vdW-DF & 3.971 & 5.377 & 73.43 & 0.2369 \\[2mm]
		optB88-vdW & 3.902 & 5.331 & 70.29 & 0.2535 \\[2mm]
		optB86b-vdW & 3.863 & 5.277 & 68.20 & 0.2494 \\[2mm]
		Expt. \cite{monteiro2017} & 3.858 & 5.264 & 67.85 & -- \\[2mm]
		Calc. \cite{lei2017} & 3.808 & 5.236 & 65.75 & -- \\
		\toprule
	\end{tabular}
\end{table}

The six independent second-order elastic constants $c_{\alpha\beta}$, calculated with different approximations for the exchange and correlation functional,  are listed in Tab. \ref{tab:cmatrix} and show good agreement when compared with the experimental data available. Elastic constants with a shear component, such as $c_{14}$ and $c_{44}$, are better predicted by the optB86b-vdW functional, evidencing the weak interaction between the adjacent layers of tellurium. However, when stronger interatomic bonds are required by the deformation, such as Ni-Te and Ni-Ni bonds, related to $c_{11}$, $c_{12}$ and $c_{33}$, the GGA-type functionals provide more accurate descriptions.

\begin{table}[t]
	\caption{Independent second order elastic constants (in GPa) calculated for trigonal NiTe$_2$.}
	\label{tab:cmatrix}
	\centering
	\begin{tabular}{lcccccc}
		\toprule
		& $c_{11}$ & $c_{12}$ & $c_{13}$ & $c_{14}$ & $c_{33}$ & $c_{44}$ \\
		\hline
		PBE & 110.8 & 38.20 & 22.90 & $-5.00$ & 45.50 & 10.20 \\[2mm]
		PBE+SOC & 113.7 & 36.60 & 27.20 & $-6.50$ & 45.70 & 11.20 \\[2mm]
		PZ+SOC & 145.5 & 54.00 & 43.30 & $-14.30$ & 76.60 & 26.80 \\[2mm]
		optB86b-vdW & 127.4 & 47.10 & 26.80 & $-9.40$ & 75.70 & 20.20 \\[2mm]
		Expt. \cite{sato1979} & 109.5 & 41.90 & -- & $-10.70$ & 52.60 & 20.40 \\[2mm]
		Calc. \cite{lei2017} & 147.6 & 50.80 & 44.10 & 7.91 & 83.90 & 17.58	\\
		\toprule
	\end{tabular}
\end{table}

The mechanical stability can be easily verified using the Mouhat and Coudert criteria \cite{mouhat2014}. The elastic anisotropy and mechanical moduli were also computed from the stiffness tensor. Surprisingly, we found that NiTe$_2$ possesses a ductile regime, favoring the strain-engineering route to manipule its low-energy excitations. These results are presented in details in App. \ref{app:mech}.

\subsection{Electronic properties}

\begin{figure*}[t]
	\centering
	\subfloat[][]{\includegraphics[width=.45\linewidth]{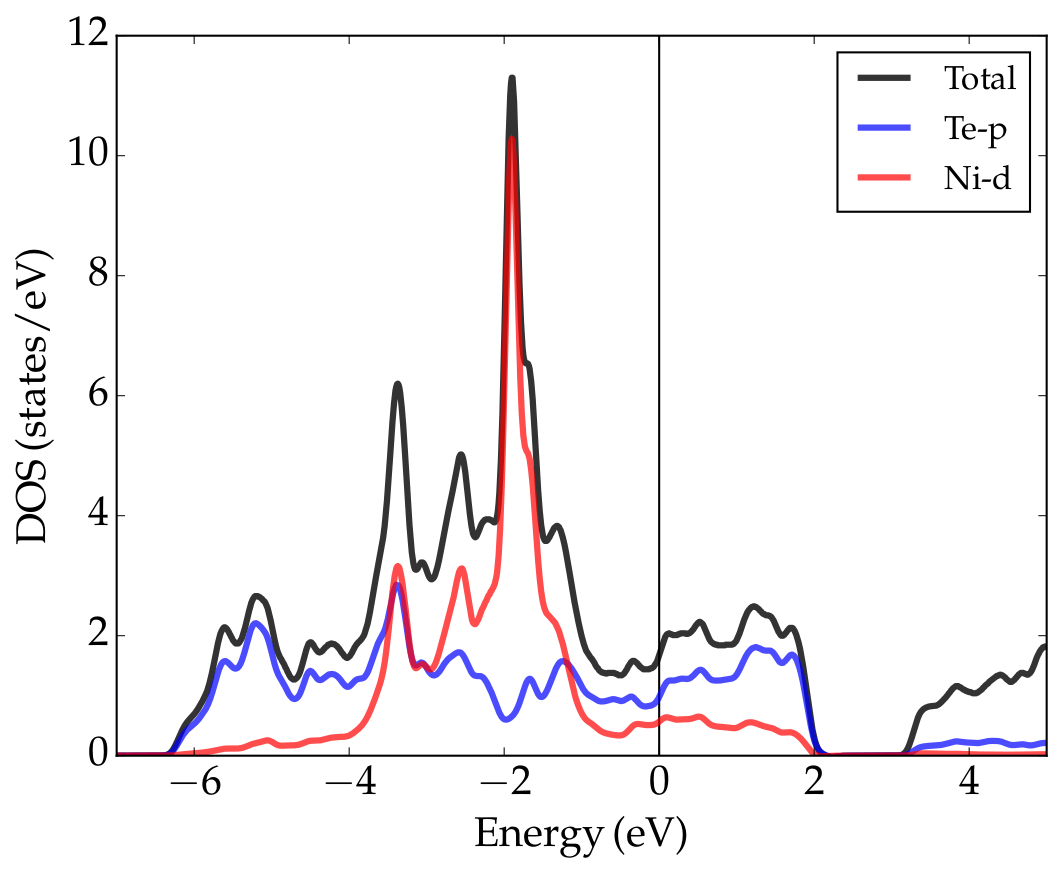}}%
	\subfloat[][]{\includegraphics[width=.51\linewidth]{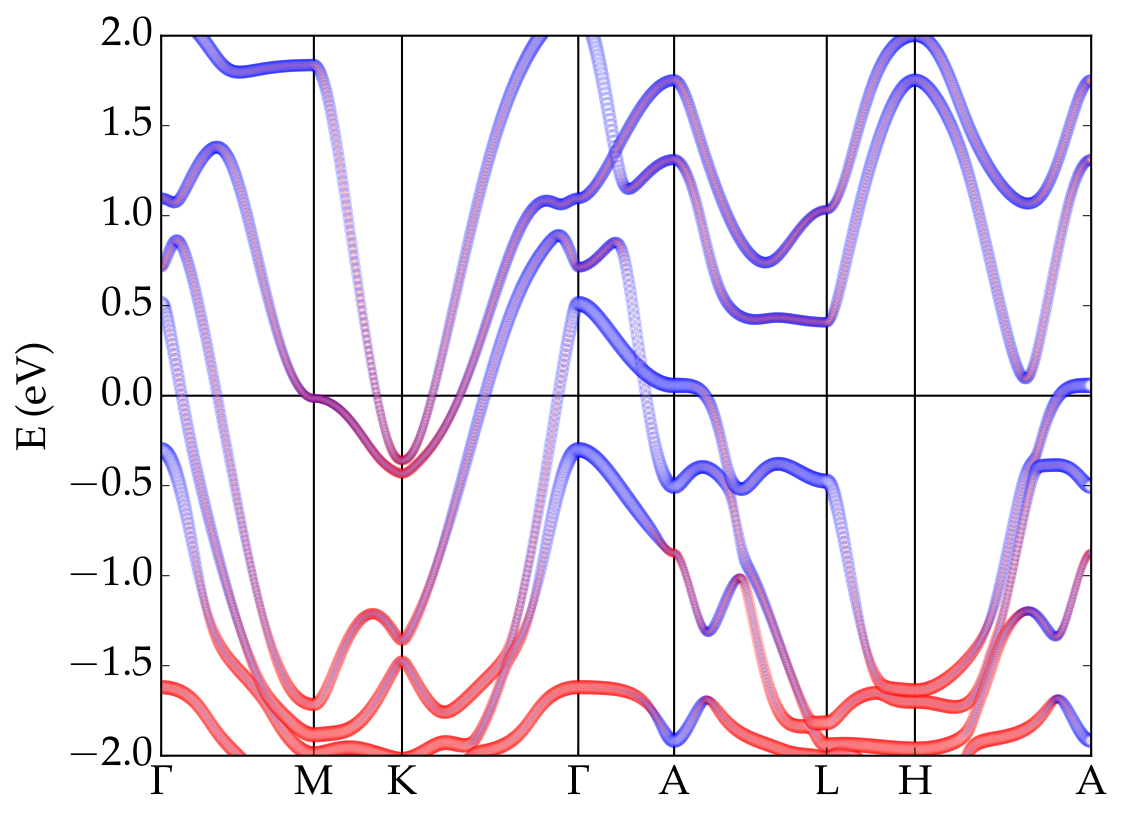}}\\
	\subfloat[][]{\includegraphics[width=.23\linewidth]{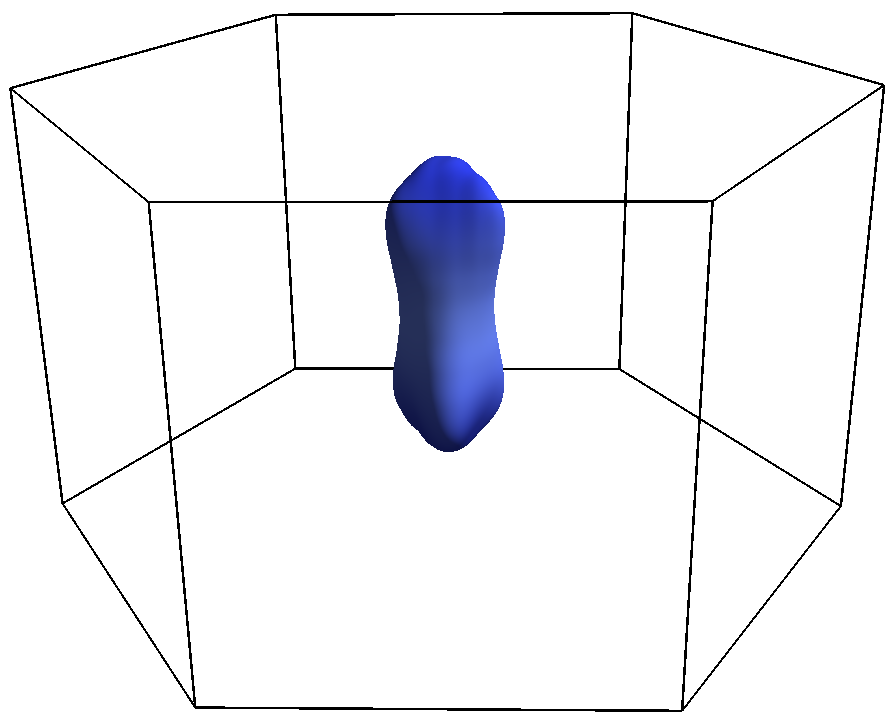}}
	\subfloat[][]{\includegraphics[width=.23\linewidth]{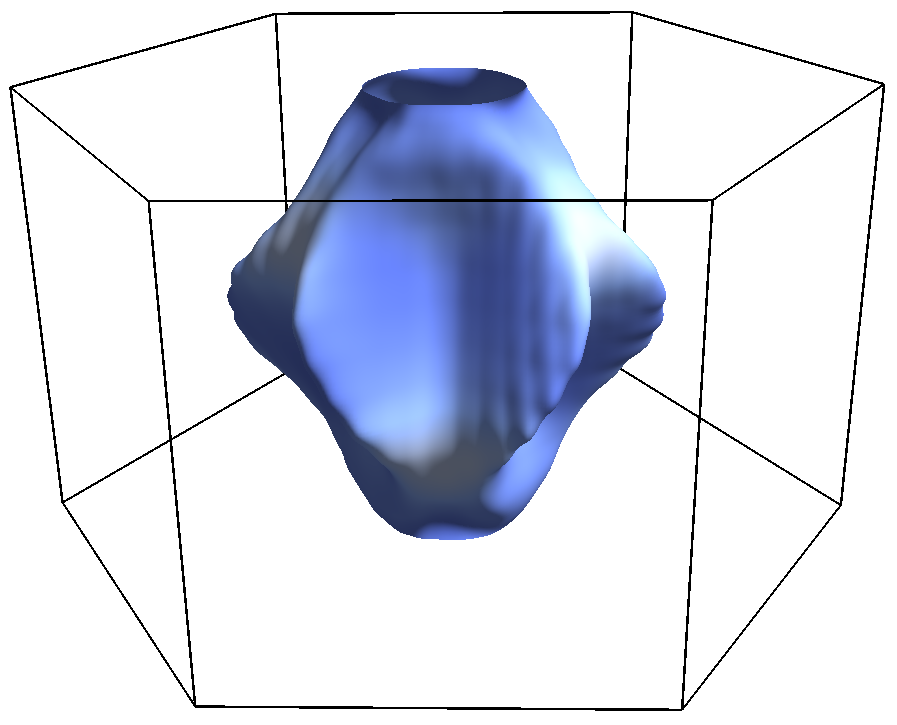}}
	\subfloat[][]{\includegraphics[width=.23\linewidth]{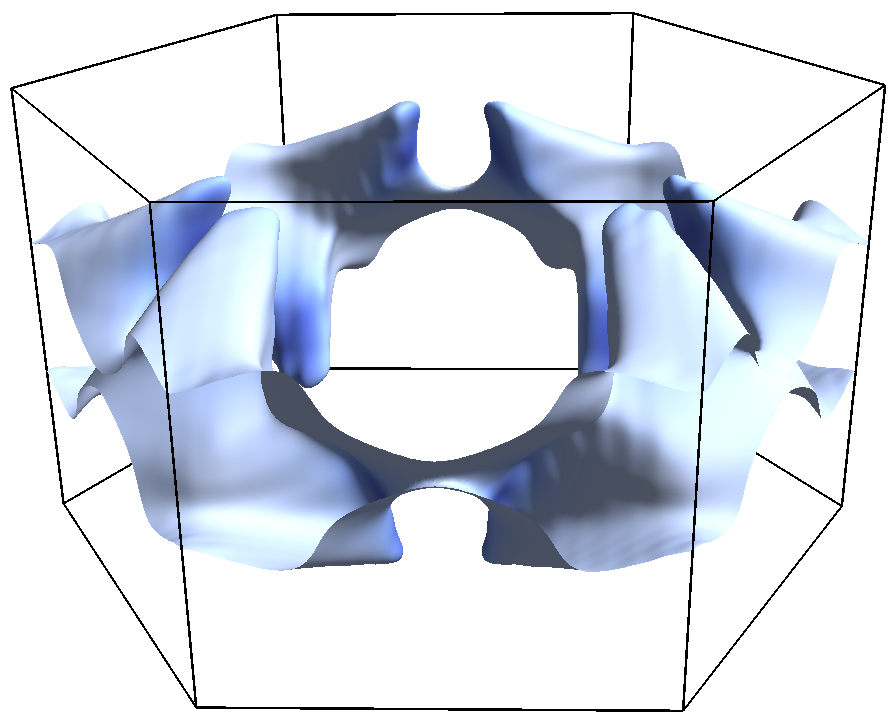}}
	\subfloat[][]{\includegraphics[width=.23\linewidth]{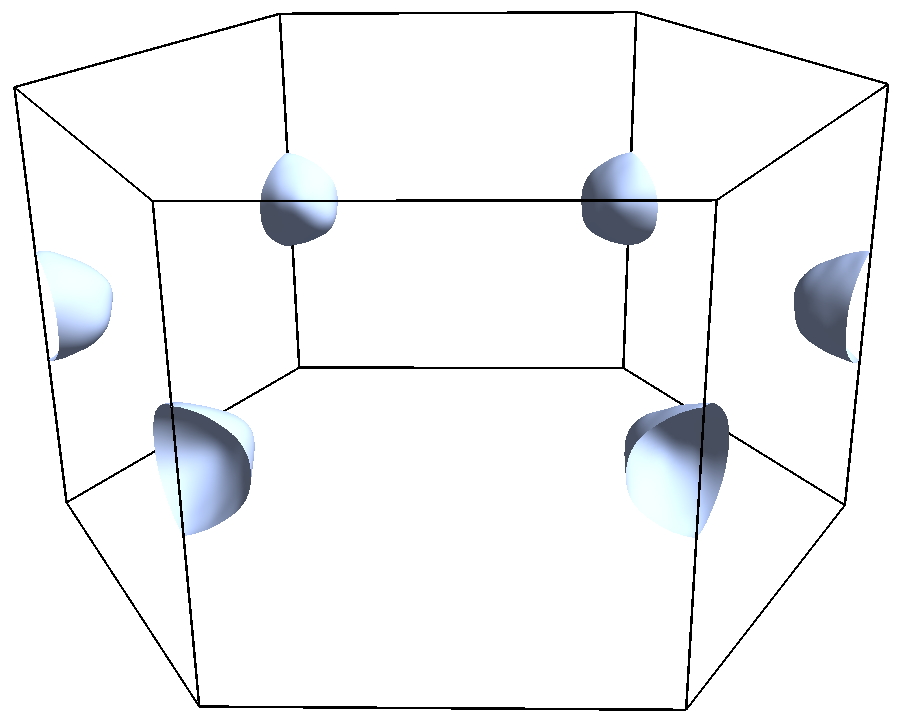}}\\
	\includegraphics[width=.46\linewidth]{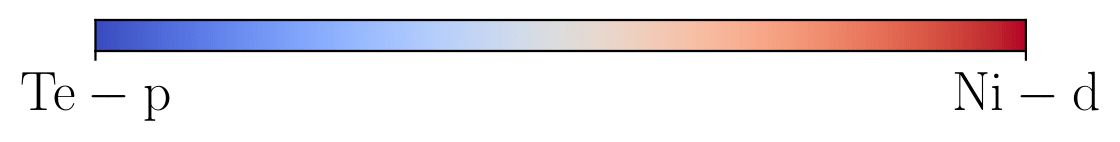}
	\caption{(a) NiTe$_2$ projected density of states and (b) electronic band-structure with SOC. (c)-(f) The four independent sheets of the Fermi surface. The color map shows the contributions of Te-5p (blue) and Ni-3d (red) manifold to the electronic wavefunction.}
	\label{fig:NiTe2-eltrn-str}
\end{figure*}

The projected density of states (DOS) of NiTe$_2$ is presented in Fig. \ref{fig:NiTe2-eltrn-str}(a). The populated Fermi level confirms the semimetal nature of the compound. The total DOS at the Fermy energy ($E_F$) is 1.67\,states/eV, with nearly 59\% of the electronic states derived from Te-$5p$ orbitals and 34\% from Ni-$3d$ manifold. Fig. \ref{fig:NiTe2-eltrn-str}(b) shows the projected electronic band-structure along path in the first Brillouin zone  shown in Fig. \ref{fig:NiTe2}.(c) There are four distinct bands crossing the Fermi level, giving rise to the four independent sheets of the Fermi surface shown in Fig. \ref{fig:NiTe2-eltrn-str}(c)-(f).

Despite the layered quasi-2-dimensional nature of NiTe$_2$, weakly coupled along the $c$ direction, the Fermi surface possesses a strong 3-dimensional character. The sheets consist of a closed (a) and an open (b) hole-pocket, and electron-pockets (c-d) around the M and K points. The hole-pockets have a strong Te-$p$ character, whereas the electron-pockets have a nearly equal contribution from Ni-$d$ and Te-$p$ states.

The valence and conduction bands touch each other at a discrete point along $\Gamma$--A, as well as the point with opposite momentum. The presence of inversion- and time-reversal symmetry ensures that each band is doubly degenerate. Therefore, the linear crossing of the valence and conduction bands originate a pair of gapless Dirac nodes (fourfold degenerate) located at $\mathbf{k}_D = (0,0,\pm0.665)$, in units of $\pi/c$. The tilted Dirac cone lies at $E_D = \unit[0.15]{eV}$. For comparison, the Pd- and Pt-based dichalcogenides host Dirac points high above the Fermi level, between 0.6 and \unit[1.2]{eV} \cite{yan2017, zhang2017, noh2017}.

\begin{figure}[t]
	\centering
	\includegraphics[width=.7\columnwidth]{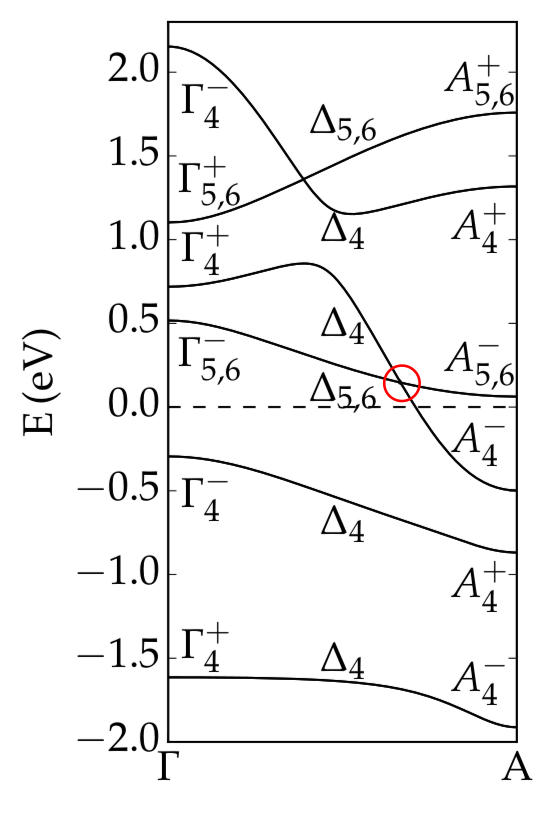}
	\caption{Detailed electronic band-structure with irreductible representations and parity analysis along the $\Gamma$--A direction. The type-II Dirac cone is circled in red.}
	\label{fig:NiTe2-irrep}
\end{figure}

The trigonal crystal-field with the strong intralayer hybridization between the Te-$p$ manifold of different sites breaks the original $p$-orbitals threefold degeneracy, resulting in bonding and anti-bonding combinations of the in-plane ($p_x,\ p_y$) and out-of-plane ($p_z$) states. The spin-orbit coupling further splits the $p$-derived electronic states due to the double group symmetry representation, including the spin degree of freedom. Since the $\Gamma$--A direction preserves the $C_3$ rotational symmetry (the system is invariant under rotations by $2\pi/3$ around the $z$ axis), the $p$-derived states will split into two distinct irreducible representations, $R_4$ and $R_{5,6}$, being $R_4$ bidimensional and $R_{5,6}$ degenerate. The irreducible representations and their parities in high-symmetry points are shown in Fig. \ref{fig:NiTe2-irrep}.

The crossing between $R_4^{\pm}$ and $R_{5,6}^\mp$ states will lead to the bulk type-II Dirac node (circled in Fig. \ref{fig:NiTe2-irrep}). This crossing is symmetrically allowed and is protected against hybridization/gap-opening mechanisms due to the $C_3$ rotational symmetry. On the other hand, the crossings between $R_4^\pm$ and $R_4^\mp$ bands are not allowed, as they both share the same symmetry and angular momentum. Therefore, due to their opposite parities, their hybridization leads to a gap with band inversion, establishing a $\mathbb{Z}_2$ invariant. This single-orbital manifold mechanism of bulk Dirac cones is widely discussed elsewhere \cite{bahramy2018, clark2019, mukherjee2020}.

Lastly, just above the Dirac point, located in close vicinity of the Fermi level, there is another band crossing giving rise to type-I Dirac fermions at $\mathbf{k}_D = (0,0,\pm0.388)$ with $R_4^{\pm}$ and $R_{5,6}^\mp$ representations and energy $E_D = \unit[1.36]{eV}$, an energy level comparable to type-II Dirac nodes in Pd- and Pt-based dichalcogenides \cite{yan2017, zhang2017, noh2017}.

\section{Strain-engineering}
\label{sec:strain}

We now turn to investigate how different strain states modify the electronic properties of NiTe$_2$. To this end, calculations were performed with three types of strain: uniaxial deformation along the [001] direction ($\mathbf{z}$-axis); biaxial deformation within the basal plane, perpendicular to the $\mathbf{z}$-direction; and an isostatic deformation. For each type, at least 6 deformations were performed, going from -5\% to +5\% with respect to the ground-state structure.

\subsection{Effects of strain in the band-structure}
\label{sec:strain-on-bands}

\begin{figure}[b]
	\centering
	\subfloat[][$\mathbf{\eta} = (-0.05,-0.05,-0.05)$]{\includegraphics[width=.5\columnwidth]{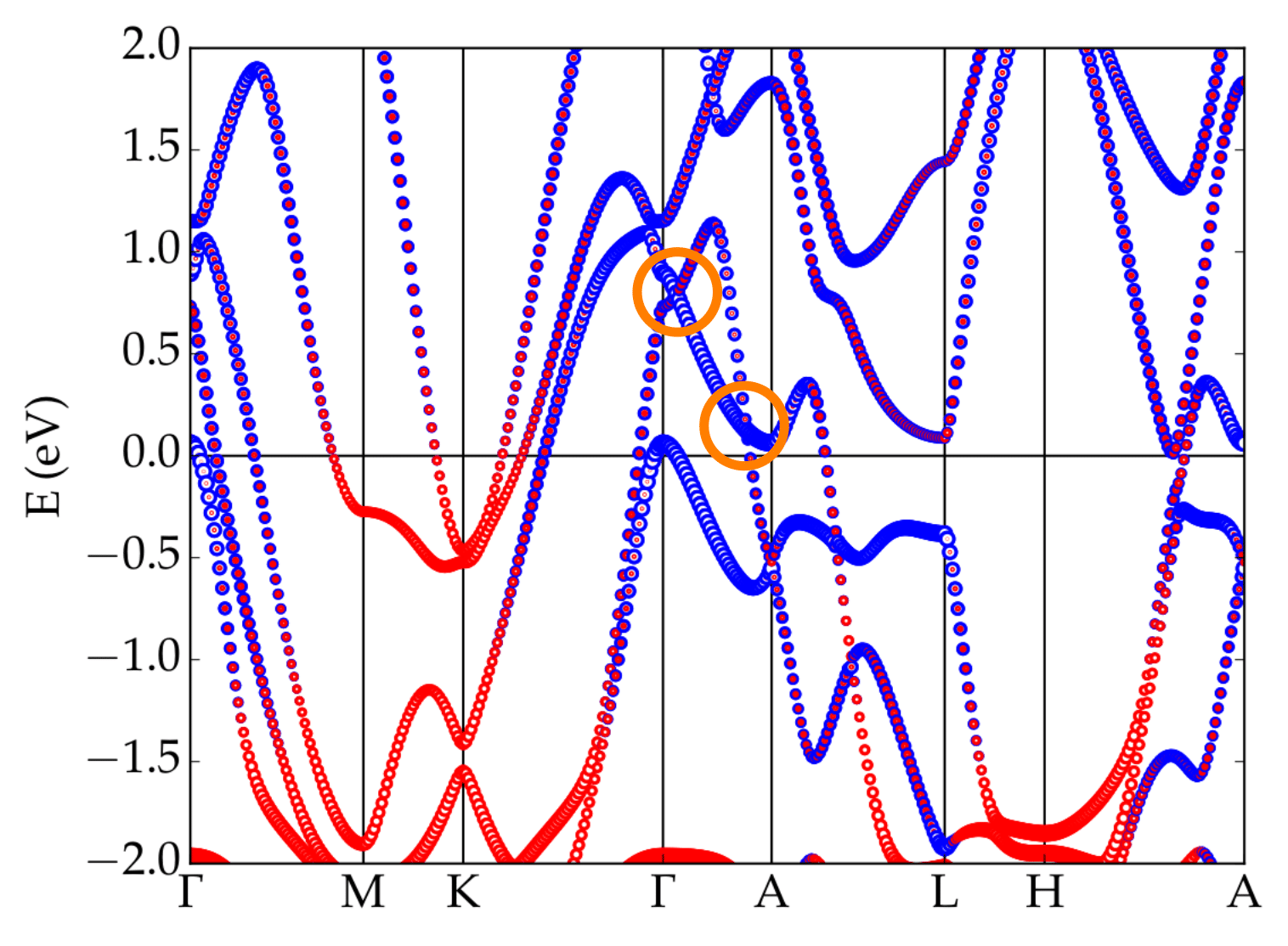}}
	\subfloat[][$\mathbf{\eta} = (0.05,0.05,0.05)$]{\includegraphics[width=.5\columnwidth]{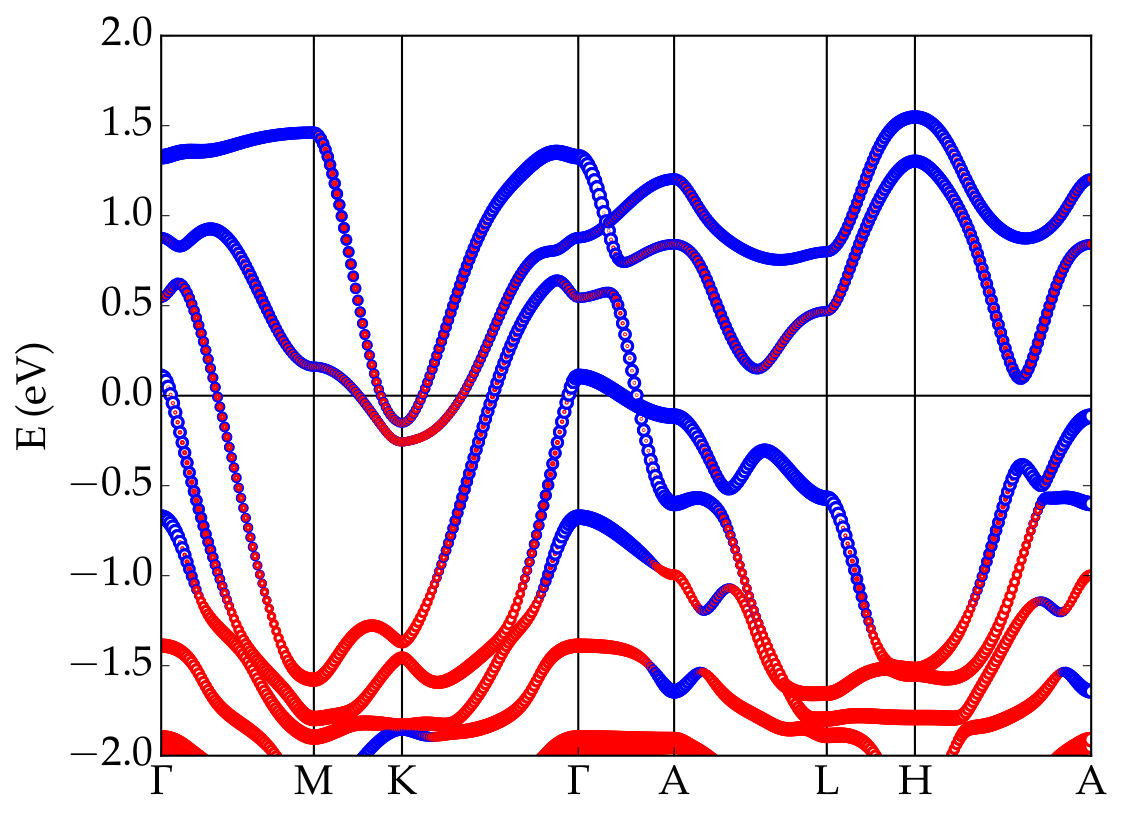}} \\
	\subfloat[][$\mathbf{\eta} = (-0.05,-0.05,0)$]{\includegraphics[width=.5\columnwidth]{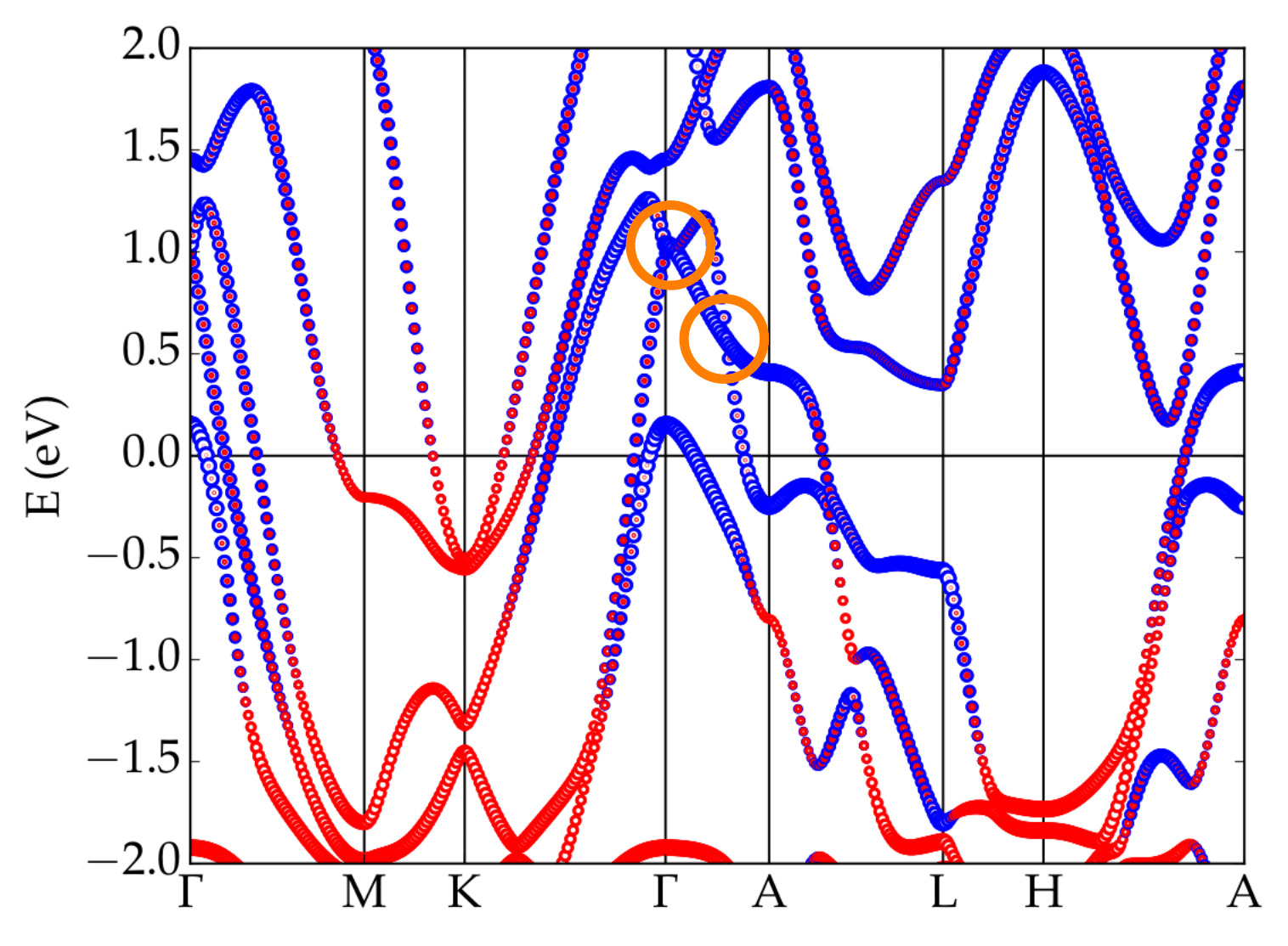}}
	\subfloat[][$\mathbf{\eta} = (0.05,0.05,0)$]{\includegraphics[width=.5\columnwidth]{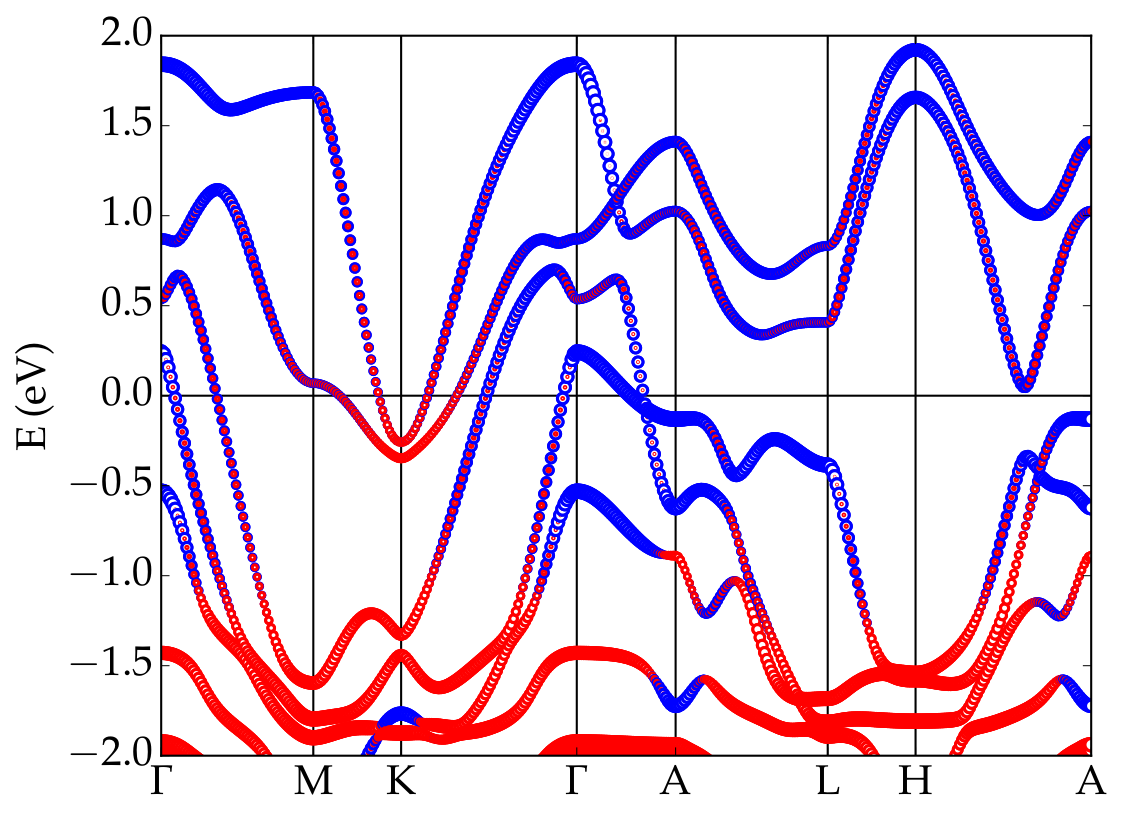}}\\
	\subfloat[][$\mathbf{\eta} = (0,0,-0.05)$]{\includegraphics[width=.5\columnwidth]{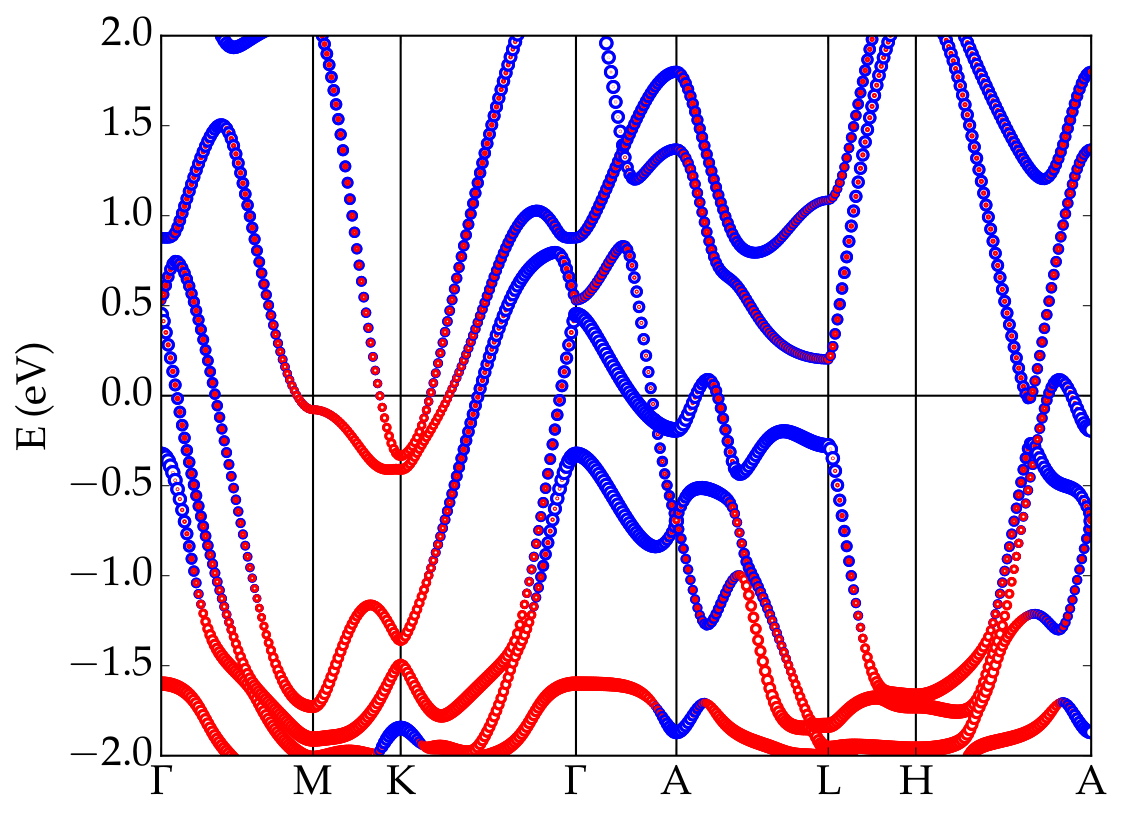}}
	\subfloat[][$\mathbf{\eta} = (0,0,0.05)$]{\includegraphics[width=.5\columnwidth]{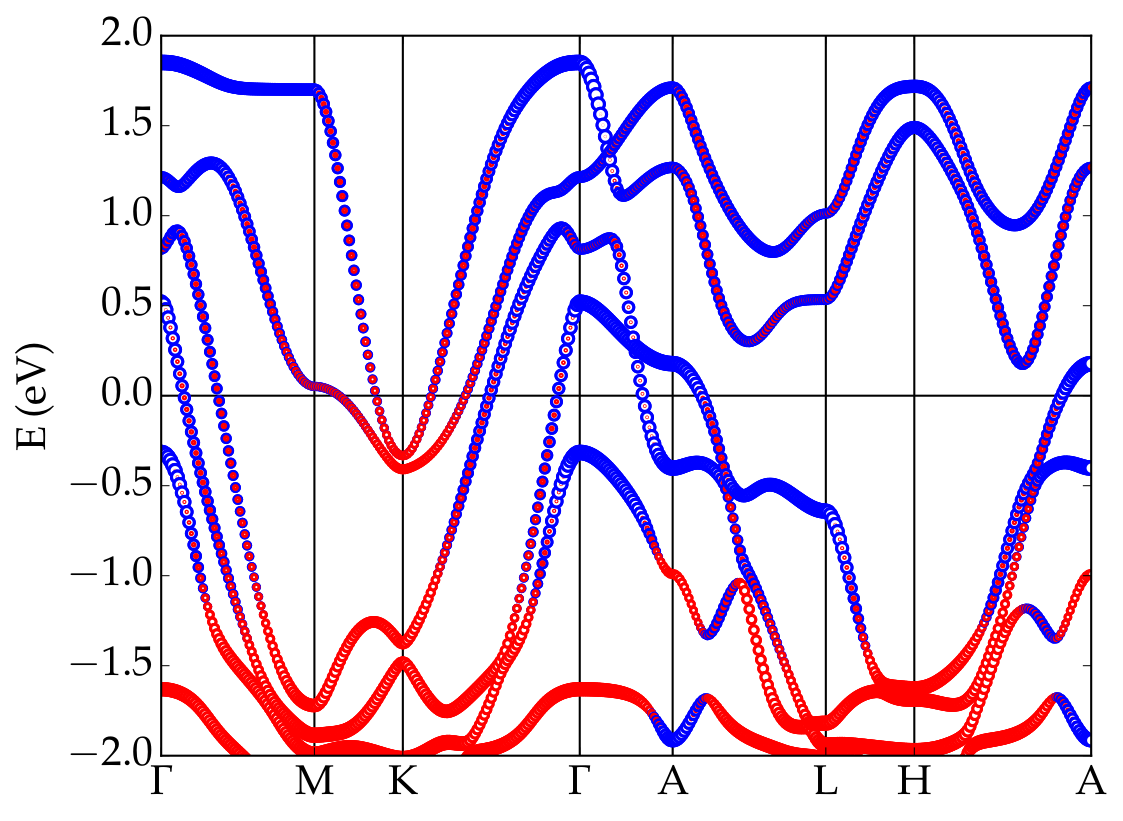}}\\
	\caption{Projected band-structure of NiTe$_2$ for some selected deformed structures. Strain states are indicated below each plot. Red points indicate the Ni-d orbital contribution and blue points, the Te-p derived states. The coexisting type-I and type-II Dirac cones in (a) and (c) are circled in orange.}
	\label{fig:NiTe2_band_strained}
\end{figure}

Fig. \ref{fig:NiTe2_band_strained} shows selected strained band-structures taking into account spin-orbit coupling effects. We observe that strain drives the hybridization of atomic orbital states and the dispersion of the bands. With compressive strain, the intralayer and interlayer couplings increase, enhacing the hybridization of electronic wavefunctions. In this way, the hopping parameters increase, culminating in bands with high effective velocities, as can be seen in Figs. \ref{fig:NiTe2_band_strained} (a), (c), and (e). Otherwise, as we move the atoms away from each other with tensile strain, the overlap between the wavefunctions decrease, resulting in bands with lower velocities, as shown in Figs. \ref{fig:NiTe2_band_strained} (b), (d), and (f). Consequently, it is possible to tune the tilt parameter of the the Dirac cone, promoting controlled changes in the anisotropic transport properties. The Dirac point energy with respect to the Fermi level is also tunable.

\begin{figure}[t]
	\centering
	\subfloat[][$\mathbf{\eta} = (-0.02,-0.02,-0.02)$]{\includegraphics[width=.5\columnwidth]{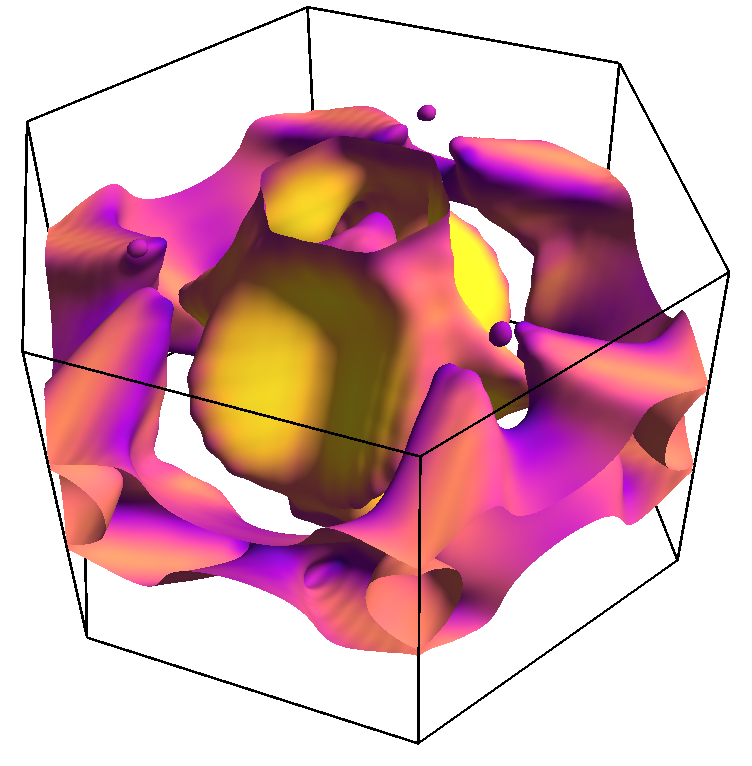}}
	\subfloat[][$\mathbf{\eta} = (0.02,0.02,0.02)$]{\includegraphics[width=.5\columnwidth]{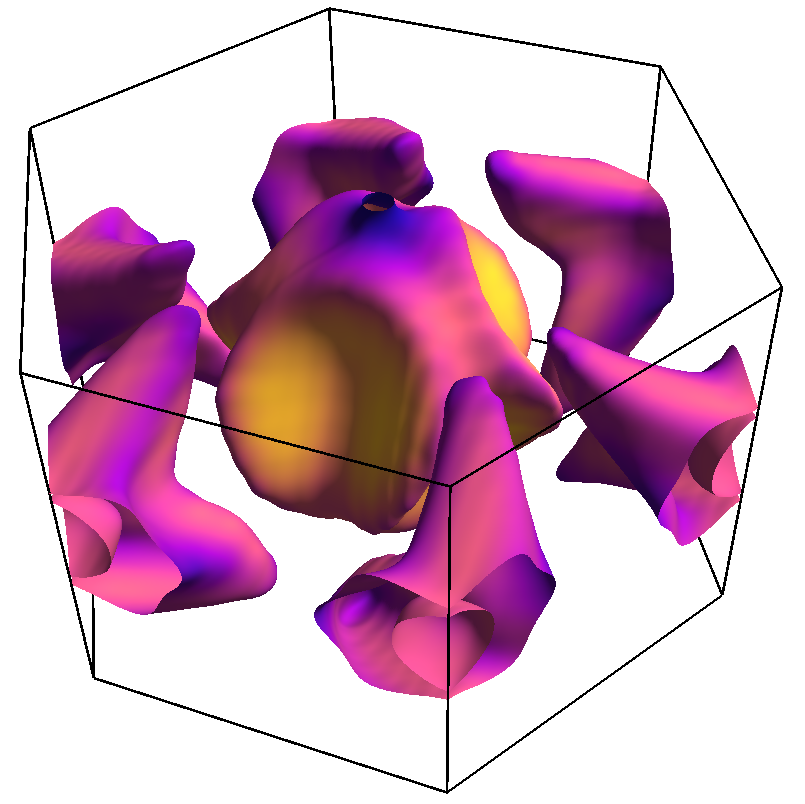}}\\
	\includegraphics[width=.8\columnwidth]{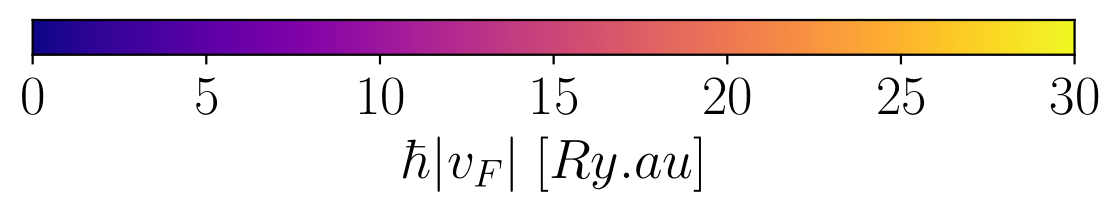}
	\caption{Fermi surfaces of NiTe$_2$ under isostatic strain. Strain states are indicated below each plot. The color map indicates the magnitude of the Fermi velocity.}
	\label{fig:NiTe2_fermi_strained_Dst01}
\end{figure}

We can also check that some non-Dirac bands crossing the Fermi level are suppressed when the structure is expanded, and extra bands become part of the Fermi surface when the structure is compressed. To illustrate this, the velocity operator projected onto the Fermi surface under isostatic deformation is shown in Fig. \ref{fig:NiTe2_fermi_strained_Dst01}. When the structure is compressed (Fig. \ref{fig:NiTe2_fermi_strained_Dst01}a), an additional branch, corresponding to the irreducible representation $R_4$, shows up in the Fermi surface along the $\Gamma$--A direction. On the other hand, by separating the adjacent layers (Fig. \ref{fig:NiTe2_fermi_strained_Dst01}b), we suppress the crossing along the A--L direction and create an intersection along $\Gamma$--A. As a result, we will have a disconnected hole-pocket and an electron-pocket surrounding the $\Gamma$-point and two electron-pockets surrounding the $K$-point, with lower Fermi velocity. As expected, the type-II Dirac point appears in the contact between the electron-pocket and the hole-pocket that are located around $\Gamma$ as we bring the chemical potential to $E=E_D$. If we lower the chemical potential further, the surfaces disconnect again and the electron-pockets will gradually decrease in size. From this perspective, the isoenergetic surfaces will evolve rapidly and undergo a sudden change in their topology due to lattice deformations, paving the way for Lifshitz transitions \cite{liu2019_2, qi2020}.

\subsection{Dynamically controlling the Dirac cone}

\begin{figure}[t]
		\centering
		\includegraphics[width=.95\columnwidth]{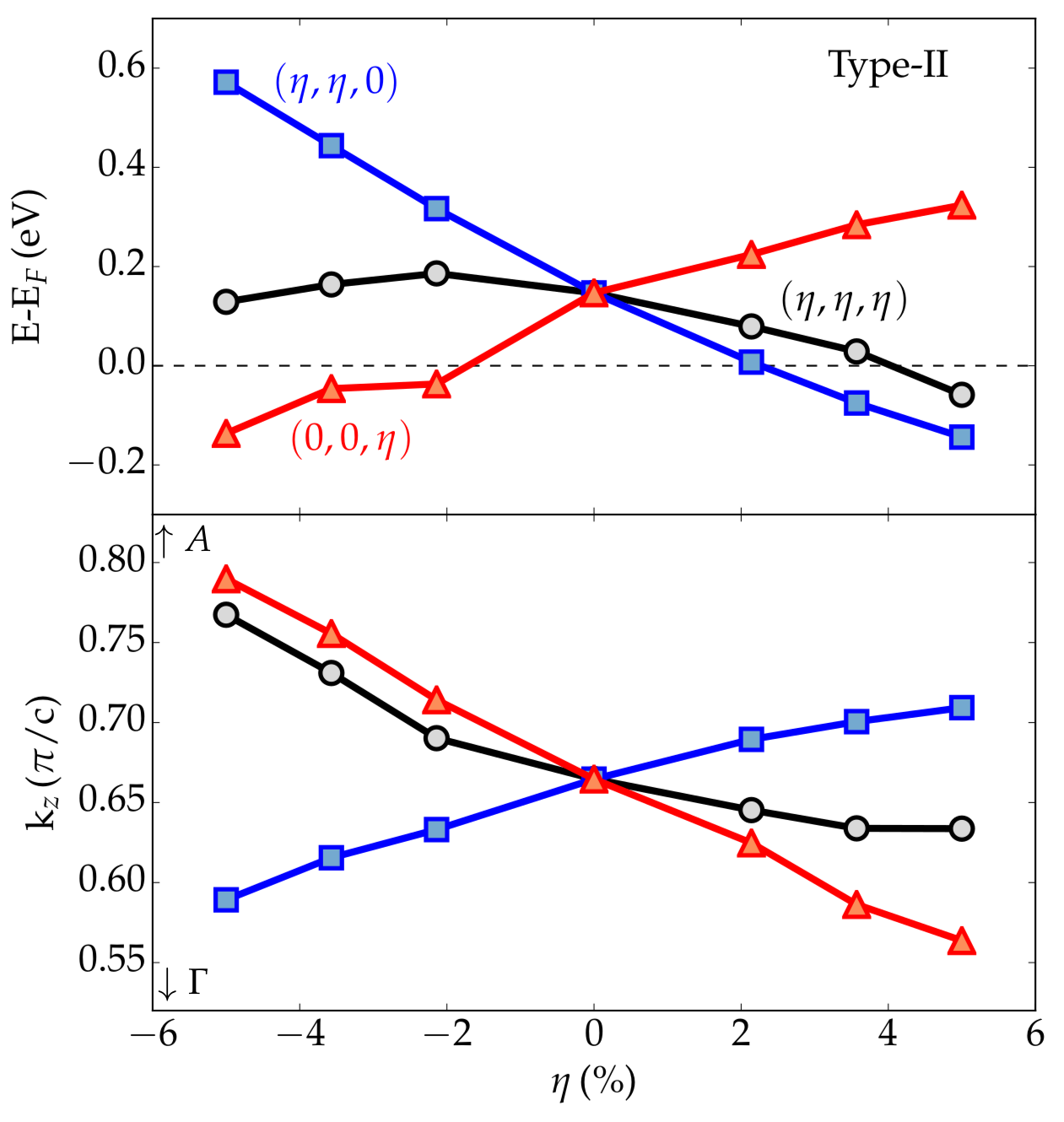}
		\caption{Type-II Dirac node energy and momentum evolution as a function of: (i) biaxial deformation (squares) within the $x$-$y$ plane $(\eta, \eta, 0)$; (ii) uniaxial strain (triangles) along the $z$-axis, $(0,0,\eta)$; and (iii) isostatic pressure (circles), $(\eta, \eta, \eta)$. A and $\Gamma$ are at $k_z=\pi/c$ and $k_z=0$, respectively.}
		\label{fig:NiTe2_DP2}
\end{figure}

In Fig. \ref{fig:NiTe2_DP2} we show the type-II Dirac node evolution under strain. For a biaxial strain, the Dirac cone moves towards the Fermi level and closer to A in $k$-space, at the border of the Brillouin zone, crossing the chemical potential at approximately $\eta = 2\%$. For a state of compression, the cone departs from the Fermi level, reaching around \unit[0.6]{eV} at $-5$\%, and approaches the center of the BZ. The opposite effect is observed for the deformation $(0,0,\eta)$. In this situation, the cone will cross E$_F $ around $-2$\%, coming close to \unit[0.4]{eV} at +5\%. Hence, for an isostatic deformation, the type-II Dirac point dynamics could be described as a combination of uniaxial and biaxial deformations. The crossing at the Fermi level occurs only close to +4\%, and the curve suggests that a compression greater than 5\% brings the cone below the Fermi energy. It is also worth noticing that, extrapolating both the Dirac node energy as well as the position in $k$-space of non-Dirac bands (see Fig. \ref{fig:NiTe2_band_strained}), there is a parameter range for which only the Dirac bands cross the Fermi level, and the node is just a few hundreds of meV below it.

\begin{figure}
	\centering
	\includegraphics[width=.95\columnwidth]{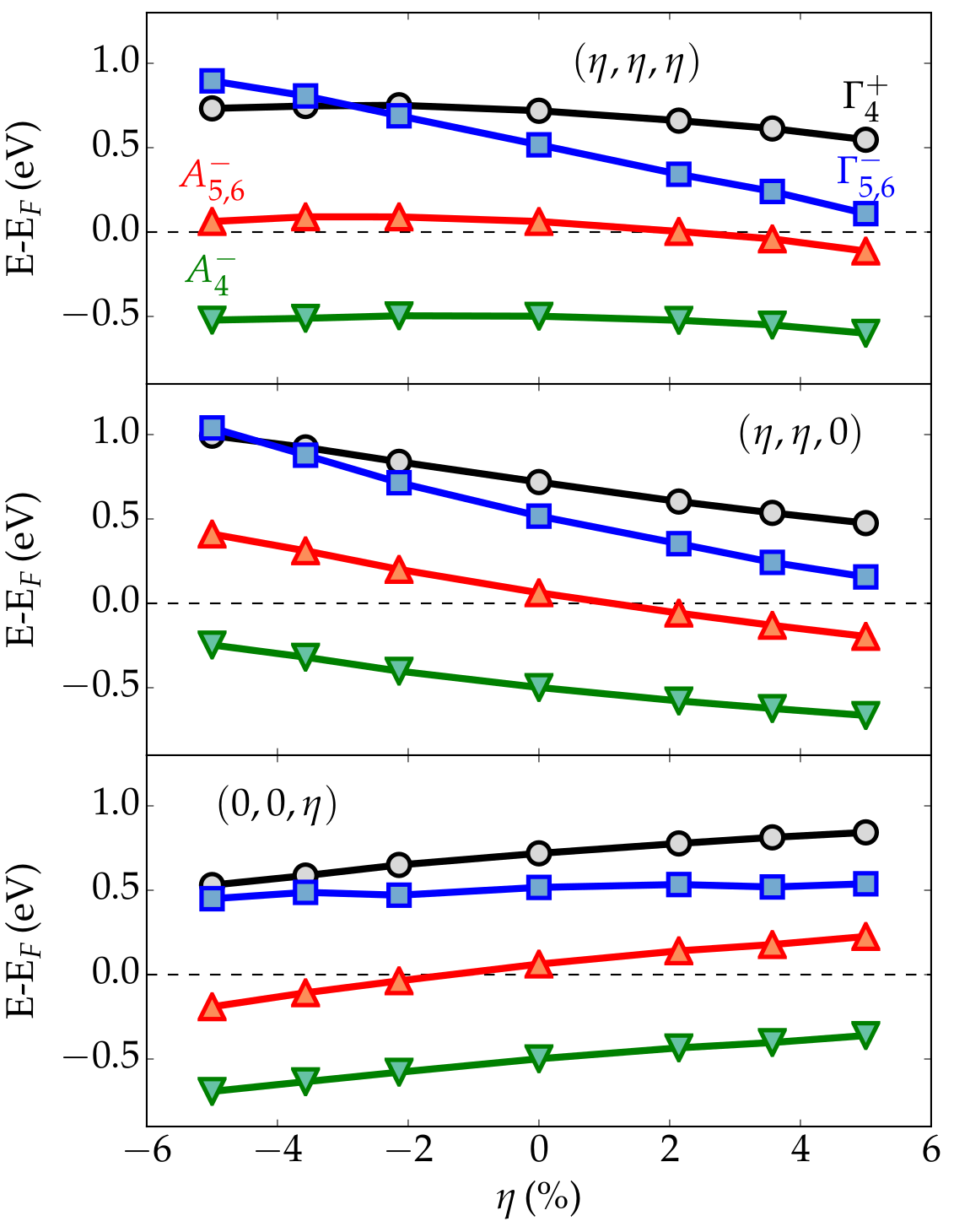}
	\caption{Energy evolution of the irreductible representations $\Gamma_{4}^+$, $\Gamma_{5,6}^-$, $A_{5,6}^-$ and $A_{4}^-$ for the deformations $(\eta,\eta,\eta)$, $(\eta,\eta,0)$ and $(0,0,\eta)$.}
	\label{fig:IR}
\end{figure}

It is also possible to observe regions in the parameter space with the coexistence of type-II and type-I Dirac cones, as shown in Fig. \ref{fig:IR}. The state $\Gamma_{4}^+$ is nearly insensitive to isostatic pressure, while $\Gamma_{5, 6}^-$ decreases monotonically. Combined, both behaviors result in a crossing between the bands with different irreducible representations. Thus, a new pair of untilted (type-I) Dirac cones is created close to $\Gamma$ for approximately $\eta=-3\% $ and $E-E_F = 0.8\,$eV (see Fig. \ref{fig:NiTe2_band_strained}a). The same effect is found for uniaxial and biaxial deformations. In the case of $(\eta, \eta, 0) $, for example, the type-I Dirac pair will form at $\eta\approx -4.2\% $ (see Fig. \ref{fig:NiTe2_band_strained}c). Thus, under strain NiTe$_2 $ harbors both type-I and type-II Dirac fermions in the same pair of energy bands.

The coexistence of type-I and type-II Dirac cones provides a route to unique and unexplored magnetoresistive and transport signatures, barely understood until now. While Dirac type-I semimetals exhibit a negative magnetoresistance in all directions \cite{xiong2015, liang2015}, the transport properties in Dirac type-II semimetals are expected to be anisotropic and present a negative magnetoresistance only in directions where the potential component of the energy spectrum is higher than the kinetic component \cite{soluyanov2015}. PdTe$_2$ \cite{xiao2017} and the family of compounds CaAgBi \cite{chen2017} are the few materials in which the coexistence of Dirac cones of type-I and type-II is expected to occur in the same pair of bands. However, the type-II Dirac node on NiTe$_2$ is much closer to the Fermi level, and, also, their momentum separation is smaller, providing a better platform to investigate interaction between quasiparticles with different pseudo-relativistic signatures.

The type-II Dirac cone strain-engineering seems an exciting route for electronic transport experiments. For example, magnetoresistance for conventional metals grows quadratically at low fields and tends to a saturation value at high fields. However, in materials where conventional charge carriers and Dirac fermions coexist and populate the Fermi surface, the magnetoresistance curve as a function of the applied magnetic field reveals an additional (and predominant) linear term \cite{abrikosov1998, abrikosov2003}. Thus, it is expected that, with the presence of Dirac cones at the Fermi level, the contributions of these quasiparticles to the magnetoresistance will be accentuated. Also, in many topological semimetals it is possible to observe a pronounced growth of the resistivity curve as temperature goes down at high fields. Enhancing the contribution from pseudo-relativistic carriers, combined with supression of carriers derived from the other non-relativistic metallic bands, it is expected that this signature will be evidenced, resulting, invariably, in a significant increase in the magnetoresistance.

\subsection{Statically controlling the Dirac cone: chemical doping}

\begin{figure*}
	\centering
	\subfloat[][]{\includegraphics[width=.75\columnwidth]{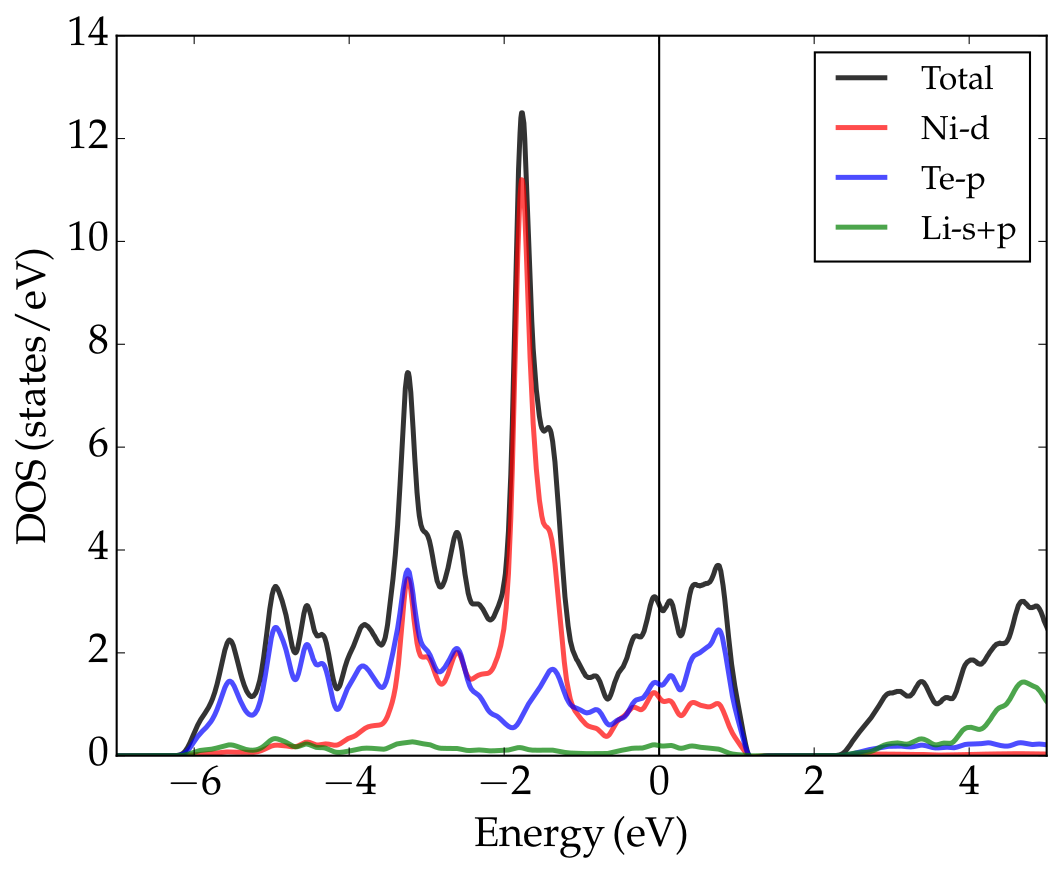}}
	\subfloat[][]{\includegraphics[width=.4\columnwidth]{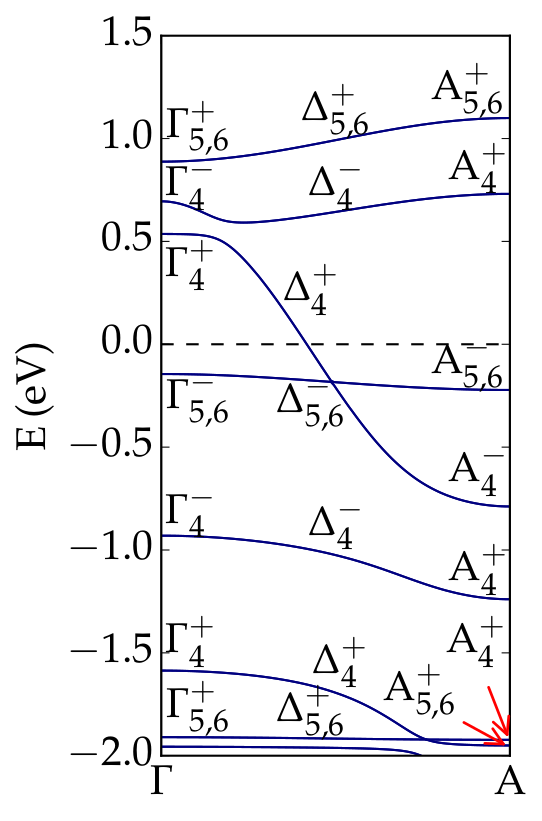}}
	\subfloat[][]{\includegraphics[width=.85\columnwidth]{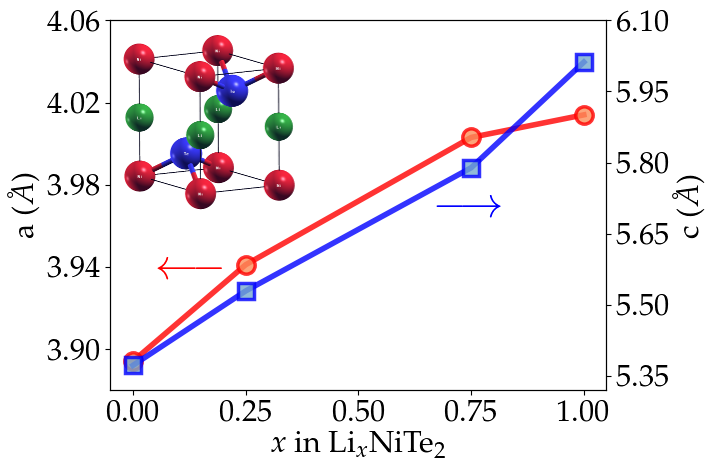}}
	\caption{(a) Electronic density of states of LiNiTe$_2$ projected onto Te-$5p$ (blue), Ni-$3d$ (red) and Li-$2s+2p$ (green) orbitals. (b) Irreducible representations and parity
symmetry of the low-energy electronic states of LiNiTe$_2$ along $\Gamma$--A. The arrows point to distinct irreducible representations in A. (c) Evolution of the $a$ (circles) and $c$ (squares) lattice parameters as a function of Li content in Li$_x$NiTe$_2$ system. The arrows in (c) indicate the corresponding axis for each curve. The inset shows the positions of the Li atoms (in green).}
	\label{fig:LiNiTe2}
\end{figure*}

Here, we show that it is possible to deform the structure without causing significant hybridization effects in the low-energy states by increasing the interlayer gap with the intercalation of alkali species. As a proof of concept, we performed first-principle calculations using the supercell method for Li$_x$NiTe$_2$ varying the Li content in the range $0\le x \le 1$. Fig. \ref{fig:LiNiTe2}(a) presents the electronic density of states of LiNiTe$_2$ projected onto Te-$5p$, Ni-$3d$ and Li-$2s+2p$ orbitals. In fact, the Li-$2s$ and Li-$2p$ states are negligible, with a very small hybridization with Te-$5p$ and Ni-$3d$ manifolds at the Fermi level. Considering a homogeneous, perfectly randomly disordered distribution of Li atoms in the lattice, all irreducible representations are conserved with Li intercalation, as shown in Fig. \ref{fig:LiNiTe2}(b), preserving the type-II Dirac cones and its topology, reproducing, therefore, the strain-modulated effects discussed in Sec. \ref{sec:strain-on-bands}. For instance, in the hypothetical situation of a full sheet into the van der Waals gap, type-II Dirac node goes below the Fermi level, at approximately $-0.2$\,{eV}. Interestingly, the $R_{5,6}^{-}$ symmetry representation is nearly flat and close to the Fermi level, opening the way for strong correlations \cite{kopnin2011,kauppila2016,roy2019}.

We can see in Fig. \ref{fig:LiNiTe2} a monotonic change in both lattice parameters as a function of Li content, reflecting the expansion of the cell in all axes. At $x = 0.25$, which is an experimentally feasible doping level \cite{morosan2006,wagner2008,morosan2010,kamitani2013,ryu2015}, we achieve 3\,\% of deformation on the $c$ axis and 1.2\,\% of deformation on the $a$ direction. Additionally, at the same compostion, Na and K atoms will promote greater deformations in the lattice as they possess a larger atomic radii than Li. Hence, it is possible to access different strain states by controlling the type and the quantity of the dopant species.

\section{Minimal effective model for type-II Dirac cones}
\label{sec:minimal-model}

With the information collected from the first-principle calculations presented in Sec. \ref{sec:properties-in-equilibrium}, we construct an effective model that describes the type-II Dirac cones. First, we note that a minimal Hamiltonian should include NiTe$_2$ discrete symmetries, namely, $C_3$ rotations, inversion, reflection along the $x$-axis and time-reversal. We restrict the Hilbert space to Te-$p$ orbitals, based on the orbital-projected band structure shown in Fig. \ref{fig:NiTe2-eltrn-str} and fix the representation with the corresponding angular momentum states:
\begin{equation}
\psi = \left(
\begin{array}{c}
J = 1/2, J_z = 1/2\\
J = 1/2, J_z = -1/2\\
J = 3/2, J_z = 3/2\\
J = 3/2, J_z = -3/2
\end{array}
\right).
\end{equation}

We then search for a family of Hamiltonians compatible with the symmetry group with Qsymm \cite{Varjas_2018}. We also restrict to a $k\cdot p$ model up to second order. The family of Hamiltonians is
\begin{align}
  \mathcal{H}(\mathbf{k}) = \epsilon(\mathbf{k}) \mathbb{1}
  + \left( \begin{array}{cccc}
  M(\mathbf{k}) & 0 & iAk_+ & Bk_+ \\
  0 & M(\mathbf{k}) & Bk_- & iAk_- \\
  -iAk_- & Bk_+ & -M(\mathbf{k}) & 0\\
  Bk_- & -iAk_+ & 0 & -M(\mathbf{k}),
  \end{array} \right),
  \label{eq:hamiltonian}
\end{align}
with
\begin{align}
  \epsilon(\mathbf{k}) &= \epsilon_0 + \epsilon_1(k_x^2 + k_y^2) + \epsilon_2 k_z^2,\\
  M(\mathbf{k}) &= M_0 + M_1(k_x^2 + k_y^2) + M_2 k_z^2,\\
  k_{\pm} &= k_x \pm i k_y.
\end{align}
Finally, fitting the DFT data, we find $M_0 = \unit[0.562]{eV},\ M_1 = \unit[-1.33]{meV \AA^2},\ M_2 = \unit[-5.10]{eV \AA^2},\ \epsilon_0 = \unit[0.873]{eV}, \epsilon_1 = \unit[-3.38]{eV \AA^2},\ \epsilon_2 = \unit[-6.58]{eV \AA^2}$ and  $A = B = \unit[3.82]{eV \AA}$.

\begin{figure}
	\includegraphics[width=\linewidth]{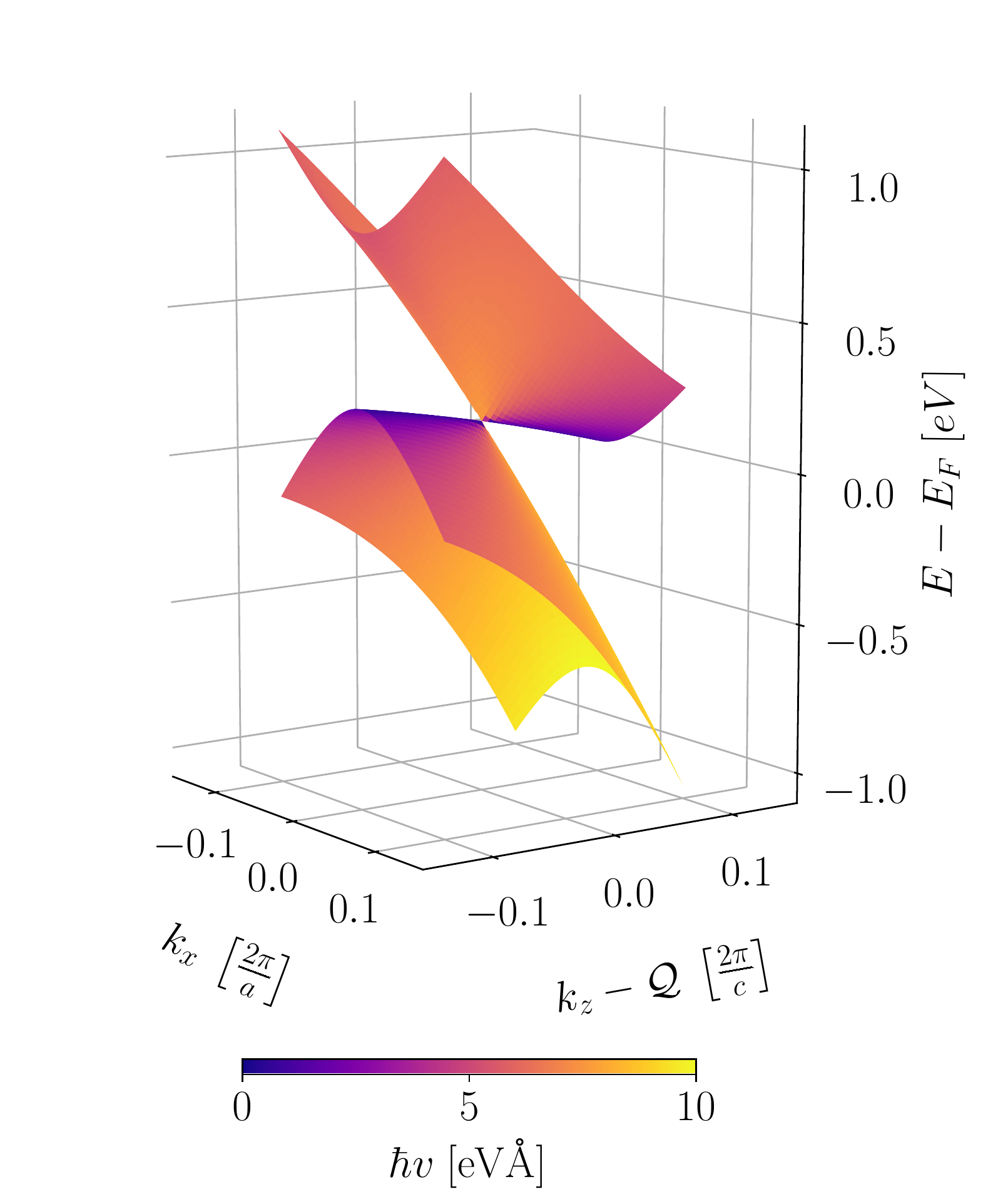}
	\caption{Dispersion around the Dirac cone obtained by diagonalizing Eq. \ref{eq:hamiltonian}. The charactetistic tilt of type-II Dirac cones is visible. The colors indicate the absolute value of the velocity at each point.}
	\label{fig:cone_3d}
\end{figure}

The dispersion around the Dirac cone obtained with the effective model is shown in Fig. \ref{fig:cone_3d}, where we clearly see the characteristic tilt of type-II Dirac cones. Furthermore, it is straightforward to check that the Dirac nodes are located at $\mathbf{k}_D = (0, 0, \pm \mathcal{Q})$, with $\mathcal{Q} = \sqrt{-M_0 / M_2}$ and at the energy $E_D = \epsilon_0 + \epsilon_2 \mathcal{Q}^2$.

Investigating the third order terms in momentum, available in the supplementary material \cite{antonio_manesco_2020_4279841}, we noticed a slight difference with respect to the description of a similar system, PtSe$_2$ \cite{PhysRevB.94.121117}. The reason is that the model used in that case is compatible with A$_3$Bi systems \cite{PhysRevB.85.195320}, which show $C_6$ rotation symmetry, whereas both NiTe$_2$ and PtSe$_2$ present $C_3$-symmetry, lacking C$_6$. The model with $C_6$ rotation symmetry is also derived in our supplementary material for the sake of comparison \cite{antonio_manesco_2020_4279841}. Moreover, we also noticed that, as expected, a gap opens at the Dirac node when $C_3$ symmetry is broken, as discussed in App. \ref{app:c3-breaking}.

We also capture the effects of strain in the type-II Dirac cones with the effective model. We build a model restricted to strain states keeping all discrete symmetries, such that Eq. \ref{eq:hamiltonian} holds, but with different parameter values. Thus, we restrict strain states to $(\eta_{x}, \eta_{y}, \eta_z)$, with $\eta_{x} = \eta_{y} = \eta_{xy}$. To recover the notion of a crystal we perform lattice regularization, and then all Hamiltonian parameters are expanded up to first order on strain (more details in App. \ref{app:lattice-regularization}). Finally, we performed a $k \cdot p$ expansion to find that the hamiltonian is modified as:
\begin{align}
	\label{eq:ham_strain}
	M_0 &\mapsto M_0 + \frac{4M_1}{\tilde{a}^2} \eta_{xy} \beta_{xy}^M + \frac{2M_2}{\tilde{c}^2} \beta_z^M \eta_z,\\
	\label{eq:ham_strain2}
	\epsilon_0 &\mapsto \epsilon_0 + \frac{4\epsilon_1}{\tilde{a}^2} \eta_{xy} \beta_{xy}^{\epsilon} + \frac{2\epsilon_2}{\tilde{c}^2} \beta_z^{\epsilon} \eta_z,\\
	\label{eq:ham_strain3}
	M_1 &\mapsto (1 - \beta_{xy}^{M} \eta_{xy}) M_1,\\
	\label{eq:ham_strain4}
	\epsilon_1 &\mapsto (1 - \beta_{xy}^{\epsilon} \eta_{xy}) \epsilon_1,\\
	M_2 &\mapsto (1 - \beta_{z}^{M} \eta_{z}) M_2,\\
	\label{eq:ham_strain5}
	\epsilon_2 &\mapsto (1 - \beta_{z}^{\epsilon} \eta_{z}) \epsilon_2,\\
	\label{eq:ham_strain6}
	A(B) &\mapsto (1 - \beta_{xy}^{A(B)} \eta_{xy}) A(B),
\end{align}
where $\tilde{a} = (1 + \eta_{xy})a$ and $\tilde{c} = (1 + \eta_z)c$ are the lattice parameters under strain and the fitted Grüneisen parameters from DFT data are $\beta_{xy}^M = -5875,\ \beta_{xy}^{\epsilon}=-0.354,\ \beta_z^M = 10.9,\ \beta_{xy}^{\epsilon} = 5.14,\ \beta_{xy}^{A(B)} = 0.083$. Moreover, the Dirac cone location in the Brillouin zone is shifted as
\begin{equation}
	\label{eq:dirac_k_shift}
	\mathcal{Q} \mapsto \mathcal{Q} \left(1 - \frac{2 M_1 \beta_{xy}^M \eta_{xy}}{M_2 \mathcal{Q}^2 \tilde{a}^2} - \frac{\beta_z^M \eta_z}{\mathcal{Q}^2 \tilde{c}^2} \right),
\end{equation}
while the Dirac node energy changes as
\begin{equation}
	\label{eq:dirac_E_shift}
	E_D \mapsto E_D + \frac{4\epsilon_1}{\tilde{a}^2} \eta_{xy} \beta_{xy}^{\epsilon} + \frac{2\epsilon_2}{\tilde{c}^2}  \beta_z^{\epsilon} \eta_z + (1 - \beta_{z}^{\epsilon} \eta_{z}) \epsilon_2 \mathcal{Q}.
\end{equation}
The Dirac cone tunability with isostatic pressure is shown in Fig. \ref{fig:dirac_tunability}, which summarizes the effects, on the position of the Dirac cone, of strain states that do not break the $C_3$ rotation symmetry. 

In the present work, we considered only uniform strain states, but it is worth to emphasize effects of non-uniform strain. Equation \ref{eq:dirac_k_shift} suggests that, in this case, a local dependency of the Dirac cone momentum, generating pseudo-Landau levels, with direct consequences to transport properties \cite{PhysRevX.6.041046}. Furthermore, when combined with electromagnetic fields, it results in a chiral anomaly \cite{PhysRevX.6.041021}. With non-uniform strain, however, Eq. \ref{eq:hamiltonian} hardly holds, since non-uniform strain will likely break some of the discrete symmetries. However, the Dirac cone will split into two Weyl cones and the observable consequences will still be present \cite{PhysRevX.6.041046, PhysRevX.6.041021}.

It is worth mentioning that the model derived here, including the effects of strain, is not restricted to NiTe$_2$, but works for any system within the same symmetry group, for example, other TMDs which are type-II Dirac semimetals, such as PtSe$_2$, PdSe$_2$ and PtTe$_2$ \cite{PhysRevB.94.121117, zhang2017, xiao2017}.

\section{Conclusions}

We presented the elastic behavior of NiTe$_2$ and its electronic structure dependency on the strain state. By analysing the electronic states' irreducible representations at high-symmetry points in the first Brillouin zone, we concluded that a type-II Dirac cone is formed by a single-orbital manifold band-inversion mechanism. Furthermore, we have shown that bulk NiTe$_2$ possesses a ductile regime, making it a candidate for electronic structure strain-engineering. Our first-principle calculations show that it is possible to tune the type-II Dirac point to the Fermi energy, making it a suitable platform for transport experiments when compared with materials of the same class \cite{PhysRevB.94.121117}. It is important to highlight that strain-engineering is achieveable in real electronic devices using piezoelectric actuators, even in mechanically delicate samples \cite{park2020}. We have also proposed a method for a static tunability with alkali metal intercalation, a process already exhaustively tested in TMDs, removing the requirement of \textit{in situ} strain control. All these effects were captured by an effective model, providing an inexpensive way for further theoretical investigations and easy comparison with experiments. In addition, the static approach with Li-doping shows the formation of dispersionless bands close to the Fermi level, favoring strong-correlation effects. Moreover, with finite strain it is possible to access hybrid type-I and type-II topological Dirac phases and promote Lifshtz transitions. Therefore, our work puts forward NiTe$_2$ as an ideal assaying platform for exploring coexisting electronic correlations and topological phenomena.

\section*{Supplementary Material}

All code and data used to prepare this manuscript is freely available on Zenodo \cite{antonio_manesco_2020_4279841}, with instructions to properly open the Python codes as Jupyter notebooks. We also added a Binder link to our Zenodo page.

\section*{Author Contributions}

P.P.F. performed and analysed the DFT calculations and wrote the first draft of the manuscript. A.L.R.M. and G.W. created the route for constructing the effective model; A.L.R.M. implemented the code and performed the calculations. T.T.D. carried out the Li-doped calculations. L.E.C. and A.J.S.M. validated the experimental considerations. L.T.F.E. supervised this project.  All co-authors revised the manuscript.

\section*{Acknowledments}

We gratefully acknowledge the financial support of the S\~ao Paulo Research Foundation (FAPESP) under Grants 2016/10167-8, 2018/10835-6, 2018/08819-2, 2019/07082-9, 2019/14359-7, 2019/05005-7, and 2020/08258-0. This study was also financed in part by the Coordena\c c\~ao de Aperfei\c coamento de Pessoal de N\' ivel Superior (CAPES) -- Brasil -- Finance Code 001. The research was carried out using high-performance computing resources made available by the Superintend\^encia de Tecnologia da Informa\c c\~ao (STI), Universidade de S\~ao Paulo.
The authors also thank Daniel Varjas, Artem Pulkin and Anton Akhmerov for fruitful discussions.

\bibliographystyle{apsrev4-1}
\bibliography{refs}

\appendix

\section{Elastic anisotropy and mechanical properties}
\label{app:mech}

\begin{table*}
	\centering
	\caption{Bulk modulus (B), shear modulus (G), Young modulus (E) and Poisson's ratio ($\nu$) for NiTe$_2 $ according to the Voigt-Heuss-Hill approximation. All values are in GPa (except dimensionless quantities).}
	\label{tab:mechanical}
	\footnotesize
	\begin{tabular}{lccccccccccccc}
		\toprule
		& $B_V$ & $B_R$ & $B_H$ & $G_V$ & $G_R$ & $G_H$ & $B/G$ & $E_V$ & $E_R$ & $E_H$ & $\nu_V$ & $\nu_R$ & $\nu_H$ \\
		\hline
		PBE & 48.34 & 38.60 & 43.47 & 23.54 & 15.60 & 19.57 & 2.22 & 60.75 & 41.24 & 51.04 &  0.29 & 0.32 & 0.30 \\[2mm]
		PBE+SOC & 50.58 & 40.56 & 45.57 & 24.34 & 16.05 & 20.19 & 2.26 & 62.93 & 42.53 & 52.79 & 0.29 & 0.33 & 0.31\\[2mm]
		PZ+SOC & 72.09 & 64.26 & 68.17 & 34.98 & 27.80 & 31.39 & 2.17 & 90.34 & 72.90 & 81.65 & 0.29 & 0.31 & 0.30 \\ [2mm]
		optB86b-vdW & 59.12 & 53.84 & 56.48 & 31.44 & 25.41 & 28.43 & 1.99 & 80.11 & 65.88 & 73.03 & 0.27 & 0.30 & 0.28 \\ [2mm]
		Calc. \cite{lei2017} & - & - & 70.12 & - & - & 28.75 & 2.44 & - & - & 50.95 & - & - & 0.32 \\
		\toprule
	\end{tabular}
\end{table*}

The second-order elastic constants $c_{\alpha\beta}$ provide valuable information about the mechanical response of a compound to a certain applied strain condition. The $c_{11}$ and $c_{33}$ constants, for instance, represent the resistance to an unixial deformation along the [100] and [001] directions, respectively, while the $c_{44}$ elastic constant is related to the resistance to a shear deformation in the ($hk0$) planes. Therefore, NiTe$_2$ has a low resistance to shearing in planes parallel to the tellurium sheets, with $c_{44} = 20.2$\,GPa, according to the optB86-vdW functional. Nevertheless, in [100] and [001] crystallographic directions we find Ni-Te and Ni-Ni bonds, offering, thereupon, greater resistance to structural changes along these directions. Thus, the constant $c_{11} = 110.8$\,GPa reflects a strong intralayer interaction, whereas the value of 45.5\,GPa for  $c_{33}$ indicates that the van der Waals gap will undergo a significant structural change when strain is applied in the [001] direction. On the other hand, nickel and tellurium will continue to interact in order to establish ionic/covalent bonds, preserving the mechanical stability and exerting some resistance.

\begin{figure}[t]
	\centering
	\subfloat[][PBE+SOC]{\includegraphics[width=.49\columnwidth]{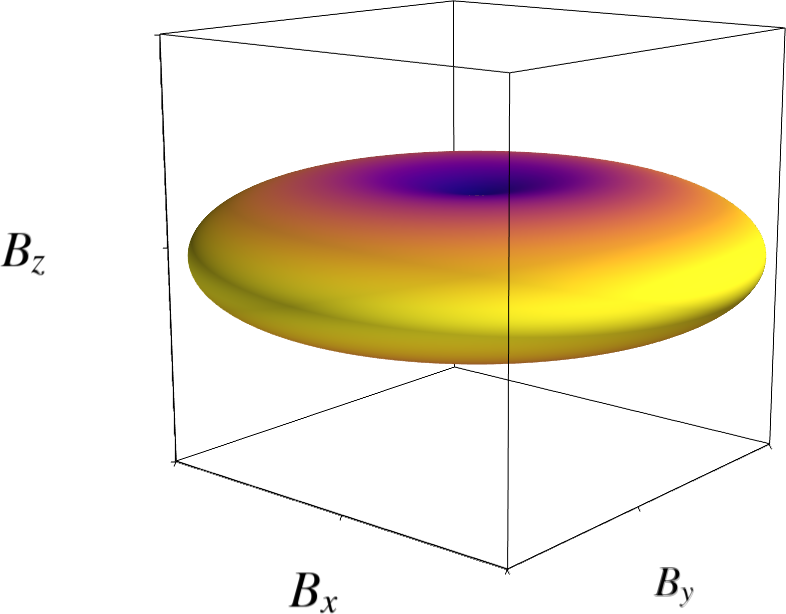}}
	\subfloat[][optB86b-vdW]{\includegraphics[width=.49\columnwidth]{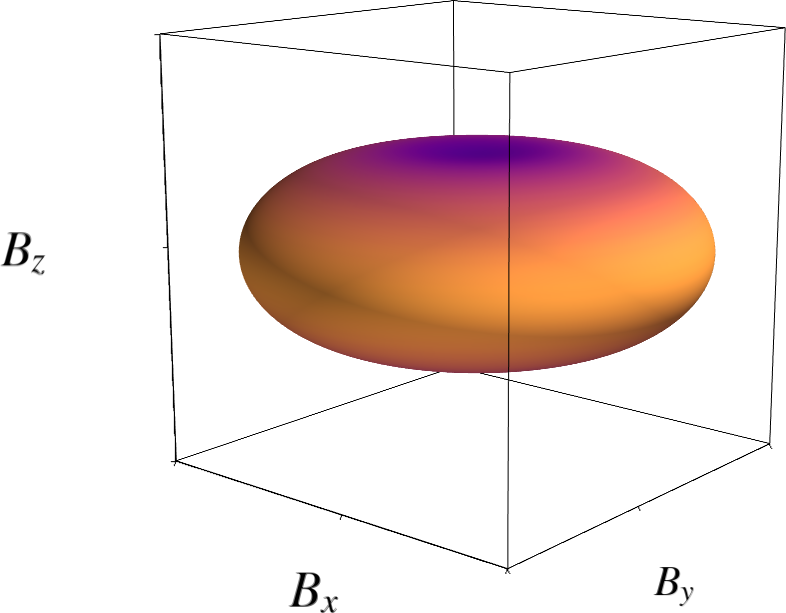}}\\
	\includegraphics[width=.9\columnwidth]{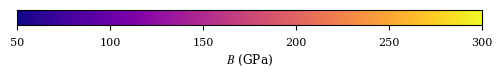}\\
	\subfloat[][PBE+SOC]{\includegraphics[width=.49\columnwidth]{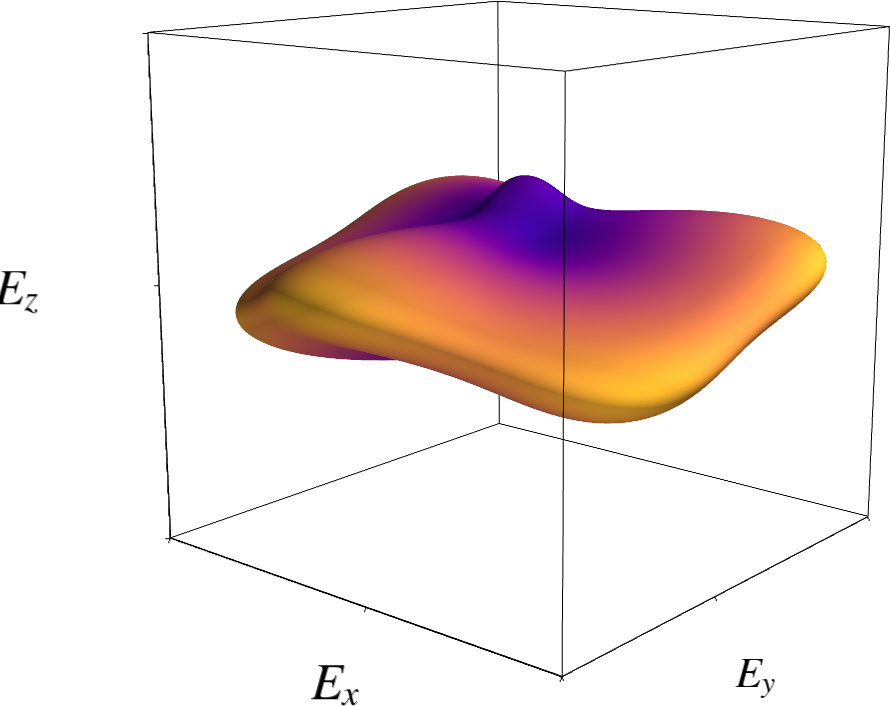}}
	\subfloat[][optB86b-vdW]{\includegraphics[width=.49\columnwidth]{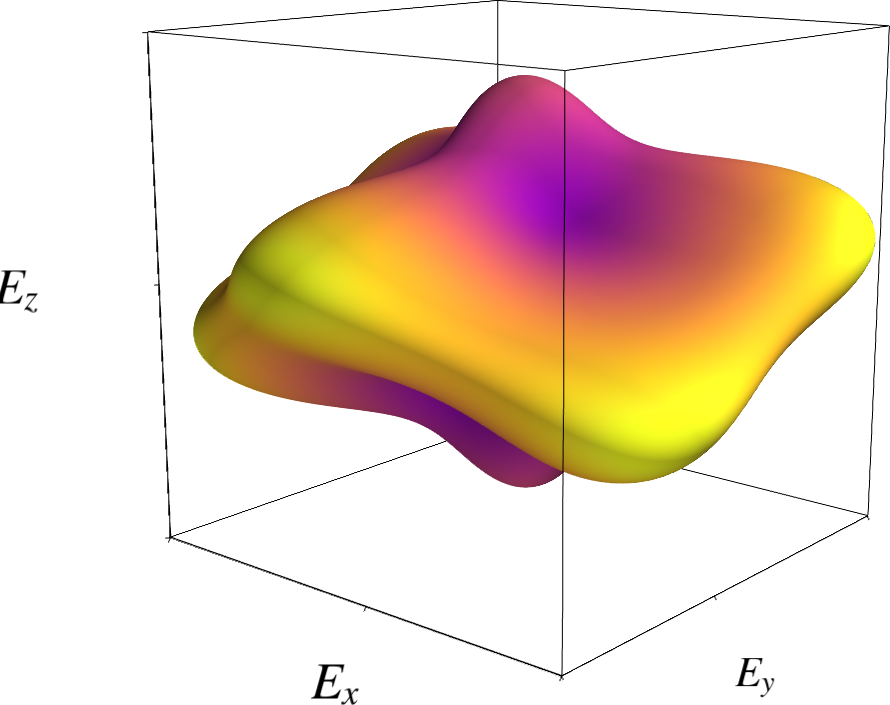}}\\
	\includegraphics[width=.9\columnwidth]{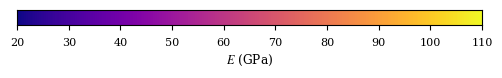}
	\caption{Directional depence of the reciprocal linear compressibility B$_c$ (a-b) and Young's modulus E (c-d) for NiTe$_2$ (in GPa) using different exchange and correlation functionals.}
	\label{fig:NiTe2_elastic}
\end{figure}

The mechanical properties within the Voigt-Reuss-Hill approximation \cite{hill1952} are shown in Tab. \ref{tab:mechanical}. It is interesting to note the B/G ratio, used as a general measure of ductility \cite{pugh1954}, for different exchange and correlation functionals. Values higher than 1.75 indicate the compound is ductile, while smaller than 1.75 indicate a brittle behavior. The calculated value, therefore, shows that NiTe$_2$, presents a good ductibility for an intermetallic compound. This assessment is consistent with a Poisson ratio higher than 0.26 \cite{chen2011}. However, the PBE+SOC approach, as expected, overestimate the ductile regime when compared to the optB86b-vdW values. This result establishes that NiTe$_2$ is a decent candidate for strain-engineering.

Knowledge of the degree of anisotropy in the single crystal elastic properties is essential to strain-engineering. The reciprocal linear compressibility ($B_c$) and Young's modulus ($E$) directional dependencies for several exchange and correlation functionals are shown in Fig. \ref{fig:NiTe2_elastic}, showing large anisotropies for $B_c$ and $E$. We observe a large resistance to elastic deformation in the [110] direction and a slight resistance along [001]. Such mechanical manifestations occur due to, as discussed based on the second-order elastic constants, the weak van der Waals interactions between adjacent Te-layers and a stronger in-plane electronic density.

\begin{figure}[h]
	\centering
	\includegraphics[width=.95\columnwidth]{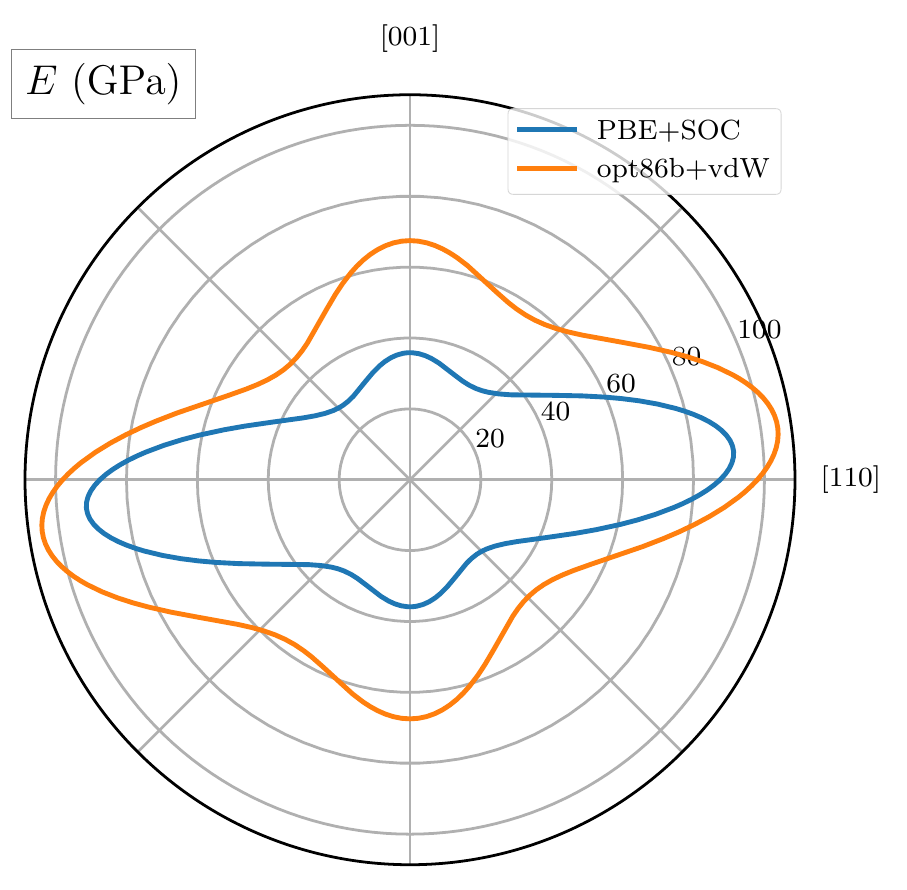}
	\caption{Polar plot of the Young's modulus $E$ (in GPa) in the (110) crystallographic plane.}
	\label{fig:NiTe2-young}
\end{figure}

It is also important to depict the changes in the elastic anisotropy profiles for different exchange and correlation functionals. Fig. \ref{fig:NiTe2-young} shows a planar projection of the Young's modulus using PBE+SOC and optB89B-vdW for directions in (110) crystallographic planes. The different mechanical resistance between the [001] and [110] directions is clear. The Young's modulus projection curve within (110) plane is visibly tilted. The origin of this elastic behavior is the opposite positions of the Te atoms in the unit cell, generating this anisotropy. Since different exchange and correlation functionals provide different force and energy minimizations to the Te atomic position degree of freedom, as well as different interatomic interactions and effective electronic densities, the net effect is a rotation of the Young's modulus projection, changing its tilting angle and its absolute values.

\section{Effects of $C_3$ symmetry breaking}
\label{app:c3-breaking}

\begin{figure}[h]
	\centering
	\includegraphics[width=.95\columnwidth]{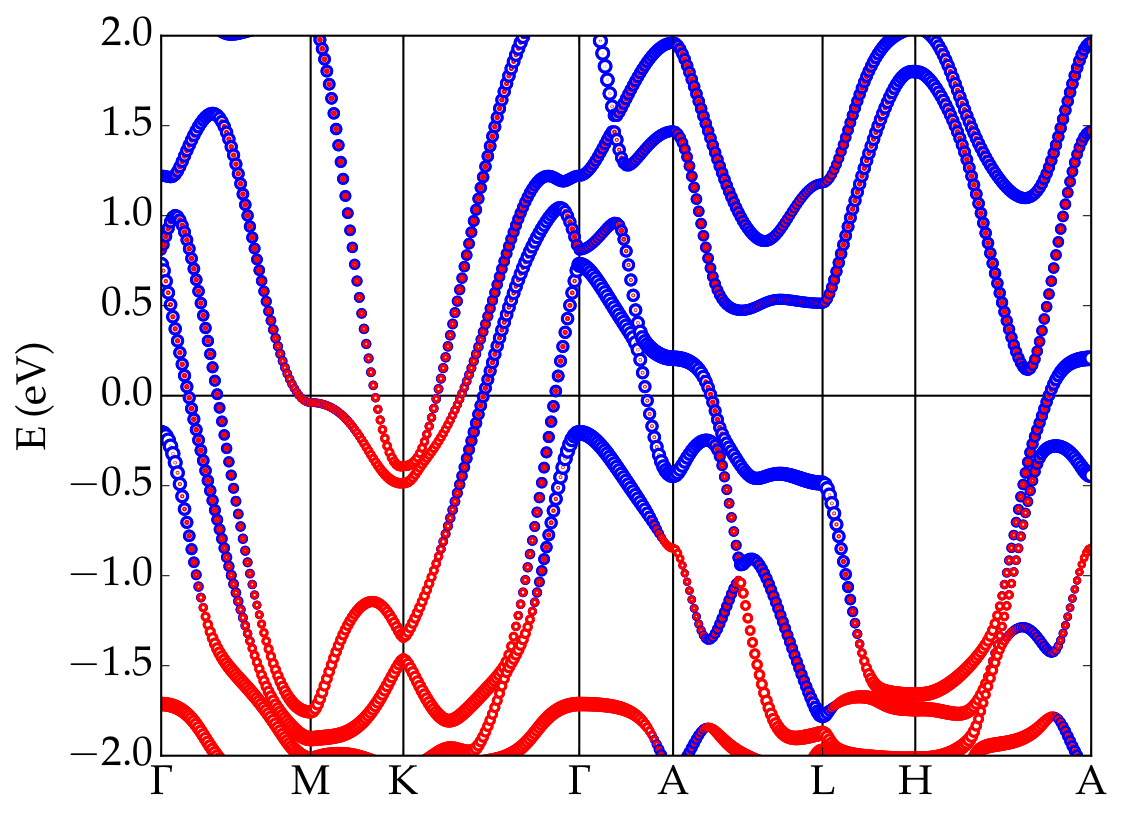}
	\caption{Projected electronic band-structure of NiTe$_2$ under in-plane uniaxial strain $\eta = (0.05, 0, 0)$. It is possible to see the gap opening at the former Dirac node position due to the C$_3$-broken symmetry. The color map shows the contribution of Te-5p (blue) and Ni-3d (red) manifold to the electronic wavefunction.}
	\label{fig:broken-C3}
\end{figure}

In Sec. \ref{sec:minimal-model} we considered only strain states that break no discrete symmetry in the system. Here, we briefly discuss the consequences of breaking the $C_3$ rotation symmetry. The procedure is straightforward: we follow the same approach as before, but we remove the constrain of 3-fold rotations. The new family of hamiltonians, then, has four aditional parameters:
\begin{align}
  \epsilon(\mathbf{k}) &\mapsto \epsilon(\mathbf{k}) + \epsilon_3 k_y k_z\ ,\quad
  M(\mathbf{k}) \mapsto M(\mathbf{k}) + M_3 k_y k_z,\\
  H(\mathbf{k}) &\mapsto H(\mathbf{k}) + \left( \begin{array}{cccc}
  0 & 0 & C k_z & iDk_z \\
  0 & 0 & -iDk_z & -Ck_z\\
  C k_z & iDk_z & 0 & 0\\
  -iDk_z & -Ck_z & 0 & 0\\
  \end{array} \right).
\end{align}
Is is noticeable, then, that there is a gap opening that is proportional to $\mathcal{Q}\sqrt{C^2 + D^2}$. We confirm that this is indeed the case by performing DFT calculations with uniaxial strain, as shown in Fig. \ref{fig:broken-C3}, that should be compared to Fig. \ref{fig:NiTe2-eltrn-str}(b).

\section{Lattice regularization and effects of strain in the Dirac cone}
\label{app:lattice-regularization}

In order to take into account the effects of strain without deriving a full tight-binding Hamiltonian, i.e., keeping the simple 4-band model presented in Sec. \ref{sec:minimal-model}, we first need to restore the idea of a lattice model. This is done via lattice regularization, implemented using the following transformations: \cite{PhysRevX.6.041021}:
\begin{align}
k_i &\mapsto \frac{1}{L_i} \sin(k_i L_i),\\
k_i^2 &\mapsto \frac{2}{L_i^2} \left[1 - \cos(k_i L_i) \right],
\end{align}
where $L_x = L_y = a$ and $L_z = c$ in the new tetragonal lattice.
Thus, the hamiltonian is now rewritten as
\begin{align}\label{eq:lattice_ham}
\mathcal{H}(\mathbf{k}) &= \tilde{\epsilon}(\mathbf{k}) \mathbb{1} + \nonumber\\
& + \left( \begin{array}{cccc}
\tilde{M}(\mathbf{k}) & 0 & i\tilde{A}_+(k_{\parallel}) + & \tilde{B}_+(k_{\parallel}) \\
0 & \tilde{M}(\mathbf{k}) & \tilde{B}_-(k_{\parallel}) & i\tilde{A}_-(k_{\parallel}) \\
-i\tilde{A}_-(k_{\parallel}) & \tilde{B}_+(k_{\parallel}) & -\tilde{M}(\mathbf{k}) & 0\\
\tilde{B}_-(k_{\parallel}) & -i\tilde{A}_+(k_{\parallel}) & 0 & -\tilde{M}(\mathbf{k})
\end{array} \right),
\end{align}
where
\begin{align}
\tilde{M}(\mathbf{k}) &= \tilde{M}_0 - \tilde{M}_1 \left[ \cos(k_z a) + \cos(k_y a) \right] - \tilde{M}_2 \cos(k_z c)\\
\tilde{\epsilon}(\mathbf{k}) &= \tilde{\epsilon}_0 - \tilde{\epsilon}_1 \left[ \cos(k_z a) + \cos(k_y a)\right] - \tilde{\epsilon}_2 \cos(k_z c)\\
\tilde{A}_{\pm} &= \frac{A}{a} \left(\sin(k_x a) \pm i \sin(k_y a)\right)\\
\tilde{B}_{\pm} &= \frac{B}{a} \left(\sin(k_x a) \pm i \sin(k_y a)\right)
\end{align}
and
\begin{align}
	\label{eq:params_latt_reg1}
	\tilde{M}_0 &= M_0 + \frac{4 M_1}{a^2} + \frac{2 M_2}{c^2}\ ,\\
	\label{eq:params_latt_reg2}
	\tilde{M}_{1} &= \frac{2 M_1}{a^2}\ ,\\
	\label{eq:params_latt_reg3}
	\tilde{M}_2 &= \frac{2 M_2}{c^2}\ ,\\
	\label{eq:params_latt_reg4}
	\tilde{\epsilon}_0 &= \epsilon_0 + \frac{4 \epsilon_1}{a^2} + \frac{2 \epsilon_2}{c^2}\ ,\\
	\label{eq:params_latt_reg5}
	\tilde{\epsilon}_{1} &= \frac{2 \epsilon_1}{a^2}\ ,\\
	\label{eq:params_latt_reg6}
	\tilde{\epsilon}_2 &= \frac{2 \epsilon_2}{c^2}.
\end{align}
It is straightforward to see that Eq. \ref{eq:hamiltonian} is recovered if we take the $k \cdot p$ expansion of Eq. \ref{eq:lattice_ham}.

Now we consider strain states $(\eta_{xy}, \eta_{xy}, \eta_{z})$, that keep all the discrete symmetries of NiTe$_2$, and, therefore, preserves the Hamiltonian form of Eq. \ref{eq:hamiltonian}. Then, we expand all parameters up to first order on strain as:
\begin{equation}
	\alpha_{i} \mapsto \alpha_{i} (1 - \eta_j \beta_j^{\alpha_i})
\end{equation}
where $\beta_{j}^i$ are the correspoding Grüneisen constants for parameter $\alpha_i$ relative to a strain state along the $j$-direction. Finally, taking a $k \cdot p$ expansion of the Hamiltonian, we obtain Eqs. \ref{eq:ham_strain} to \ref{eq:dirac_E_shift}.

\end{document}